\documentclass[pre,aps,showpacs,onecolumn,amssymb,amsmath,amstext,amsfont,noshowkeys, ]{revtex4}
\usepackage{graphicx}
\usepackage{theorem}
\usepackage{wrapfig}
\usepackage{subfigure}
\usepackage{dcolumn}
\usepackage{array}
\usepackage{color}
\usepackage{bm}
\usepackage{epsfig}
\usepackage{epstopdf}
\usepackage{hyperref}
\usepackage{changebar}
\usepackage{enumerate}
\usepackage[noprefix]{nomencl}
\makenomenclature
\usepackage{latexsym,amsmath,amsfonts,amscd,subfigure}
\newtheorem{remark}{Remark}
\numberwithin{remark}{section}
\newcommand{\PreserveBackslash}[1]{\let\temp=\\#1\let\\=\temp}
\newcolumntype{C}[1]{>{\PreserveBackslash\centering}p{#1}}
\newcolumntype{R}[1]{>{\PreserveBackslash\raggedleft}p{#1}}
\newcolumntype{L}[1]{>{\PreserveBackslash\raggedright}p{#1}}

\makeatletter
\newcommand\figcaption{\def\@captype{figure}\caption}
\newcommand\tabcaption{\def\@captype{table}\caption}

\newcommand{\Rmnum}[1]{\expandafter\@slowromancap\romannumeral #1@}
\makeatother

\begin{document}

\title{Smoothed Dissipative Particle Dynamics model for
  mesoscopic multiphase flows in the presence of thermal fluctuations}
\author{Huan Lei$^1$, Nathan A. Baker$^1$, Lei Wu$^2$, 
  Gregory K. Schenter$^1$, Christopher J. Mundy$^1$, and Alexandre M. Tartakovsky$^1$}
\affiliation{$^1$ Pacific Northwest National Laboratory, Richland, WA 99354, USA,\\
$^2$ LMAM and School of Mathematical Sciences,
Peking University, Beijing 100871 China
}

\date{\today}

\begin{abstract}
Thermal fluctuations cause perturbations of fluid-fluid interfaces and highly nonlinear hydrodynamics
in multiphase flows. 
In this work, we develop a novel multiphase smoothed dissipative particle dynamics model. This model 
accounts for both bulk hydrodynamics and interfacial
fluctuations. Interfacial surface tension is modeled by imposing a pairwise
force between SDPD particles.
We show that the relationship between the model parameters and surface
tension, previously derived under the assumption of zero
thermal fluctuation, is accurate for fluid systems at low temperature but
overestimates the surface tension for intermediate and large thermal fluctuations.
To analyze the effect of thermal fluctuations on surface tension, we
construct a coarse-grained Euler lattice model based on the mean field
theory and derive a semi-analytical formula to directly relate
the surface tension to model parameters for a wide range of temperatures and model resolutions. We demonstrate that the
present method correctly models
the dynamic processes, such as bubble coalescence and
capillary spectra across the interface.
\end{abstract}


\maketitle


\section{Introduction}
\label{sec:introduction}
Thermal fluctuations originating from molecular interactions can 
profoundly affect the behavior of multiphase fluid systems,
resulting in emergent phenomena reflected on the hydrodynamic length scale.
Consistent coupling of the molecular and hydrodynamic scales is
at the heart of mesoscale framework development.
At the fluid-fluid interface, capillary waves generated by
thermal fluctuations result in stochastic and highly nonlinear interfacial
dynamics~\cite{OrtizdeZarate2004}. This dynamics plays an important role in many
physical and biological processes, such as spreading of nano
droplets \cite{Davidovitch:2005gu}, breakup of nano-jets \cite{Moseler:2000kn},
Rayleigh-Taylor instabilities \cite{Kadau_PNAS_2007}, and protein
mobility within membranes \cite{Quemeneur_PNAS_2014}. Numerical modeling of such
processes must accurately account for
fluid momentum transport in bulk and interfacial dynamics
under thermal fluctuations.

Traditionally, thermally driven mesoscale flows are described by  so-called fluctuating hydrodynamics (FHD) (also known as Landau-Lifshitz-Navier-Stokes
(LLNS)) equations \cite{Landau1987}. These equations
extend the hydrodynamic Navier-Stokes (NS) description by adding  spatiotemporal
delta-function-correlated random stress in the NS equations with the stress covariance determined from the
fluctuation-dissipation theorem.
Several numerical methods have been developed to solve LLNS equations, including
finite difference  \cite{Atzberger_Peskin_JCP_2007, Atzberger_JCP_2011,
Voulgarakis_Chu_JCP_2009}, finite volume  \cite{Serrano_Espanol_PRE_2001,
Bell_Garcia_2007,Donev_Vanden_2010,Donev_Bell_PRL_2011}, and Lattice-Boltzmann
methods \cite{Ladd_PRL_1993}. Extending these grid-based methods to
mesoscale multiphase and/or multicomponent flows requires
coupling FHD with additional thermodynamic equations or appropriate
boundary conditions at the interface between two fluids
\cite{Espanol_Thieulot_JCP_2003, Barry_Chu_JCP_2011}.
Specifically, the interface between two fluids
yields the FHD equations highly nonlinear, and it is nontrivial to numerically
solve them with grid-based methods.
Recently, grid-based methods have been used to solve coupled LLNS-free energy
equations to model multiphase mesoscale flows (i.e., liquid and gas
phases of the same fluid). For example, Shang \textit{et al.}~\cite{Barry_Chu_JCP_2011}
solved the FHD equations coupled with the Ginzburg-Landau free energy model by using a  
heuristic correction for the observed unphysical negative density
fluctuations across the interface. Donev \textit{et al.}~\cite{Chaudhri_Donev_PRE_2014}
solved the FHD equations coupled with a free energy model based
on van der Waals equations. In this work, the surface tension is imposed through
Korteweg stress, which requires further parameter calibration and
can be sensitive to spatial discretization.

In this paper, we present a rescaled Pairwise-Force
Smoothed Dissipative Particle Dynamics (rPF-SDPD) method for mesoscale multicomponent flows.
The rPF-SDPD model combines the Pairwise-Force Smoothed
Particle Hydrodynamics (PF-SPH) method  for multiphase multicomponent flows in the absence of thermal fluctuations \cite{Tartakovsky2005, Tart-PFSPH}
 with the Smoothed Dissipative Particle Dynamics (SDPD)
\cite{Espanol_Revenga_PRE_2003, Grmela1997, Ottinger1997} method.
The major difference between PF-SPH and
the present rPF-SDPD model is that rPF-SDPD accounts for the effect of
interfacial thermal fluctuations on the interfacial surface tension and, therefore, can
be used to study mesoscale multicomponent flows.
The SDPD model can be derived from the
Smoothed Particle Hydrodynamics method (SPH) \cite{Gingold1977,Lucy1977,Monaghan2005}
by adding random forces, satisfying the fluctuation-dissipation theorem. Both SPH and
SDPD are fully Lagrangian particle methods. In both methods, the fluid domain
is discretized with a set of points (also refereed to as fluid particles).
Due to their Lagrangian nature, the SDPD and SPH methods 
are well suited for multicomponent
flow simulations because they do not require any interface tracking schemes to evolve
the interface between different fluid components. In these methods, each fluid is
represented by its own set of particles with positions advected by fluid velocities.

The SPH method has been extensively used for multicomponent and multiphase simulations.
There are two main SPH approaches for imposing surface tension at  fluid-fluid interfaces that can be extended to SDPD. The first approach is based on the Continuum
Surface Force (CSF) method  \cite{Brackbill1992,Hu2006} and requires estimating
 the local normal vectors and curvatures at the interface by means of a color
function \cite{Morris2000, Hu2006, Hu_JCP_2009}. 
Accurate
estimation of these variables requires sufficient resolution at the interface,
i.e., the radii of the largest curvature corresponding to interfacial roughness
should be much larger than the SPH smoothing parameter $h$, which plays the
same role as grid size in grid-based methods. 
It has been demonstrated that the CSF-SDPD method yields accurate results for
macroscopic multiphase flows where thermally induced interface oscillations are not pronounced  \cite{Hu2006}. For problems with large front oscillations, the effect of interfacial roughness
on the CSF-SDPD accuracy requires
additional investigations. 

The second approach is the PF-SPH
method, where pairwise molecular-like forces are added into the SPH momentum
conservation equation to produce surface tension at the fluid-fluid interface
\cite{Tartakovsky2005, Tart-PFSPH}. Unlike the CSF-based SPH method, the PF-SPH
method does not require estimates of the normal and curvature of the interface.

In this paper, we focus on moderate thermal fluctuations
and their effects on the oscillations of the interface between fluids.
Starting with the PF-SPH method, developed for low-temperature smooth
interfaces (radii of curvature much greater than $h$), we analyze the effect of
thermal-fluctuation-induced interface roughness by constructing a coarse-grained
lattice model based on a mean field theory. The coarse-grained model enables us
to extract a universal scaling relationship between surface tension and the thermal
fluctuations for various model resolutions (i.e., $h$) and construct
a semi-analytical relationship between the surface tension and model parameters.
We demonstrate that the numerical values of the surface tension imposed by the
proposed method agree well with the theoretical predictions based on the 
correct rescaled formulation. We also show that the structure factor 
of the perturbed interface correctly scales with the wave number. 

In this work, we distinguish between macroscopic (in the absence of thermal fluctuations) and mesoscopic (in the presence of fluctuations) surface tensions and use the following notation:

\begin{itemize}
\item{$\sigma$} -- {mesoscopic  surface tension across a nearly flat interface with local roughness 
induced by fluctuations due to the thermal energy $k_BT$}%
\item{$\sigma_0$} -- {macroscopic surface tension across flat and smooth interface 
(without local roughness)}
\item{$\tilde{\sigma}(R, k_BT)$} -- {mesoscopic surface tension across the interface with curvature 
  radii R and local roughness induced by fluctuations due to the thermal energy $k_BT$}%
\item{$\sigma^N$} -- {numerical value of surface tension obtained from a direct simulation}%
\item{$\sigma^F$} -- {surface tension obtained from numerical fitting to the 
scaling relationship.}%
\end{itemize}

The paper is organized as follows. 
Section~\ref{sec:model} introduces
the governing equations of the multiphase flow and their discrete
SDPD counterparts. In Section~\ref{sec:MFT}, we show that the ``zero-thermal fluctuations''
relationship between the surface tension and force parameters overestimates
the surface tension of mesoscale fluids in the presence of thermal fluctuations.
To accurately account for the effect of thermal fluctuations, we introduce a
coarse-grained lattice model based on the mean field theory.
We establish a semi-analytical relationship between
the temperature-dependent surface tension and model parameters through proper scaling among
different model resolutions. In Section~\ref{sec:Numeric}, we demonstrate that the present method
yields consistent thermodynamic properties and further show that the present method
captures the correct dynamic processes in multiphase flow, such as bubble
coalescence dynamics and  capillary wave spectra
of an interface with and without external gravity field.
Conclusions are given in Section~\ref{sec:discussion}.

\section{Numerical model}
\label{sec:model}
\subsection{Governing equations}
We consider the flow of $\alpha$ and $\beta$ fluid components, occupying the domains
$\Omega_\alpha(t)$ and $\Omega_\beta(t)$, respectively, with a sharp boundary
$\Gamma(t) = \Omega_\alpha(t) \cap \Omega_\beta(t)$ separating the two fluids. We assume that at the mesoscale, flow of these fluids can be described by the isothermal stochastic Navier-Stokes equations \cite{OrtizdeZarate2004}, including the continuity equation
\begin{equation}
\frac{D \rho_l }{D t}=-{\rho_l}\left( {\boldsymbol{\nabla} \cdot \boldsymbol{v}}_l \right),  \quad \mathbf{x} \in \Omega_l, \quad l = \alpha, \beta
\label{Eq-Cont}
\end{equation}
and the momentum conservation equation
\begin{equation}
\frac{D\boldsymbol{v}_l}{Dt}=-\frac{1}{\rho_l}\boldsymbol{\nabla} P_l + \frac{1}{\rho_l}\boldsymbol{\nabla} \cdot \boldsymbol{\tau}_l+\boldsymbol{g}
+ \frac{1}{\rho_l} \boldsymbol{\nabla} \cdot \boldsymbol{s}_l, \quad \mathbf{x} \in \Omega_l, \quad l = \alpha, \beta.
\label{linear-momentum}
\end{equation}
Here $D/Dt = \partial/ \partial t + \boldsymbol{v}_l \cdot \boldsymbol{\nabla}$ is the total derivative;
$\rho_l$, $\boldsymbol{v}_l$, and $P_l$  are the density, velocity, and pressure; and $\boldsymbol{g}$ is the  body force. The components of the viscous stress $\boldsymbol{\tau}_l$ are given by 
\begin{equation}
\begin{aligned}
\tau^{ik}_l =\ &\mu_l \left(\frac{\partial v^i_l}{\partial x^k} + \frac{\partial v^k_l}{\partial x^i} \right),
\label{eq:stresssimple}
\end{aligned}
\end{equation}
where $\mu_l$ is the (shear) viscosity ($l=\alpha, \beta$) and the bulk viscosity of the $l$-fluid component is assumed to be equal to $\frac23 \mu_l$.

Fluctuations in velocity are caused by the random stress tensor
\begin{equation}
\boldsymbol{s}_l = \gamma_l \boldsymbol{\xi},
\end{equation}
where $\boldsymbol{\xi}$ is a random symmetric tensor (which components are random Gaussian variables), and $\gamma_l$ is the strength of the noise.
The random stress is related to the viscous stress by the fluctuation-dissipation theorem \cite{Landau1987}. For incompressible and low-compressible fluids, the covariance of the stress components is:
\begin{equation}
\overline{{s^{in}_l(\mathbf{x}_1,t_1)s^{jm}(\mathbf{x}_2,t_2)}}=
\gamma^2_l \delta(\mathbf{x}_1-\mathbf{x}_2)\delta(t_1-t_2)
\hspace{0.5cm}
\gamma^2_l = 2 \mu_l k_B T \delta^{ij}\delta^{nm},
\label{eq:randomstress_squared}
\end{equation}
 where $k_B$ is the Boltzmann constant, $T$ denotes the temperature, $\delta(z)$ is the Dirac delta function, and $\delta^{ij}$ is the Kronecker delta function.

For generality, we treat the fluids as compressible and prescribe an equation of state $P_l = f(\rho_l)$ for each phase to close Eqs. (\ref{Eq-Cont}) and (\ref{linear-momentum}). Eqs.~(\ref{Eq-Cont}) and (\ref{linear-momentum}) are subject to the no-slip  boundary conditions for the fluid velocity at the fluid-solid boundary
\begin{equation}
v_{l,n} = 0\hspace{0.5cm}\text{and}\hspace{0.5cm} v_{l,\tau}=0,\quad l=\alpha,\beta,
\end{equation}
and the dynamic Young-Laplace boundary condition for pressure and velocity at the fluid-fluid-interface
\begin{equation}\label{Young-Laplace}
(P_\alpha-P_\beta)\mathbf{n}= (\boldsymbol{\tau}_\alpha-\boldsymbol{\tau}_\beta)\cdot\mathbf{n} +\kappa \sigma \mathbf{n},\quad\mathbf{x}\in\Gamma,
\end{equation}
where $v_{l,n}$ and $v_{l,\tau}$ are the normal and tangent components of velocity; $\kappa$ is the curvature of the interface; and $\sigma$ is the surface tension between $\alpha$ and $\beta$ fluids. The normal vector {$\bf n$} is pointed away from the non-wetting phase. In this work, we are interested in the dynamics of fluid-fluid interfaces, so we will only consider cases not involving fluid-fluid-solid interfaces. Otherwise, the contact angle at the fluid-fluid-solid interface would also need to be prescribed.
Eqs.~(\ref{Eq-Cont}) and (\ref{linear-momentum}) are subject to the initial conditions
\begin{equation}
\mathbf{v}_l (\mathbf{x},t=0) =\mathbf{v}_l^0 (\mathbf{x}),\quad \Omega_l (t=0)= \Omega^0_l,\quad\rho_l(\mathbf{x},t)=\rho_l^0,\quad l=\alpha,\beta.
\end{equation}

\subsection{Smoothed Dissipative Particle Hydrodynamics}

In the SDPD method, the computational domains $\Omega_\alpha$ and $\Omega_\beta$
are discretized with $N_\alpha$ and $N_\beta$ points (usually referred to as
particles) with initial positions $\mathbf{r}_i^0$. It is convenient, but not necessary, to initially put particles on a Cartesian mesh with grid size $\Delta$ discretizing the domain $\Omega=\Omega_\alpha \cup \Omega_\beta$. The particles within domains $\Omega_\alpha^0$ and $\Omega_\beta^0$ can then be labeled as $\alpha$ and $\beta$ particles, respectively, and the particles are assigned the viscosities of the corresponding fluids. The mass of particle $i$ in domain $\Omega_l$ is set to $m_i = \rho^0_l \Delta^d$, and the initial particle density of particle $i$ is $n_i^0 =n_{eq} =\Delta^{-d}$, where $d$ is the number of spatial dimensions.  Eq.~(\ref{linear-momentum}) is approximated as
\begin{eqnarray}\label{SDPD-momentum}
\frac{D \mathbf{r}_i}{D t} = \mathbf{v}_{i},\quad m_i \frac{D \mathbf{v}_{i}}{Dt}  =
\sum_{j=1}^{N}\left( \mathbf{F}_{ij}^P + \mathbf{F}^{visc}_{ij} +  \mathbf{F}_{ij}^S  \right)+\mathbf{F}_i^b,
\end{eqnarray}
where $N=N_\alpha + N_\beta$ and the summation is over all particles,
\begin{equation}
\mathbf{F}_{ij}^P
=-\left(\frac{P_j}{n_{j}^2}+\frac{P_i}{n_{i}^2}\right) \frac{\mathbf{r}_{ij}}{r_{ij}}\frac{ d W(r_{ij},h)}{d r_{ij}},
\end{equation}
\begin{equation}\label{visc-force}
\mathbf{F}^{visc}_{ij}= \frac{5 (\mu_i + \mu_j)}{6 n_i n_j}  \frac{1}{r_{ij}}\frac{d W(r_{ij},h)}{d r_{ij}}
\left[\mathbf{v}_{ij} + \left( \mathbf{v}_{ij} \cdot \frac {\mathbf{r}_{ij}}{r_{ij}} \right) \frac {\mathbf{r}_{ij}}{r_{ij}} \right],
\end{equation}
and $\mathbf{F}_{ij}^S$ is
\begin{equation}\label{eq:sdpd}
\mathbf{F}_{ij}^{S} =  B_{ij} \frac{\mathbf{r}_{ij} }{r_{ij}} d \tilde{\boldsymbol{\mathcal{W}}}_{ij},
\end{equation}
where ${\bf r}_{ij} = {\bf r}_i - {\bf r}_j $ and $r_{ij} = |{\bf r}_{ij}| $. In Eq. (\ref{eq:sdpd}),
\begin{equation}
B_{ij}=\sqrt{ -\frac{20}{6} k_B T \frac{(\mu_i+\mu_j)}{n_i n_j} \frac{1}{r_{ij}} \frac{d W(r_{ij},h)}{d r_{ij}} },
\end{equation}
$d \tilde{\boldsymbol{\mathcal{W}}}_{ij} = ( d \boldsymbol{\mathcal{W}}_{ij} +  \boldsymbol{\mathcal{W}}_{ij}^T)$, and $ \boldsymbol{\mathcal{W}}_{ij}$ is the matrix of independent increments of the Wiener process.

The random number $\tilde{\zeta}^l_{ij}=\tilde{\zeta}^l_{ji}$ has a Gaussian distribution with zero mean and unit variance, and superscript $l$ denotes $l$-component of vectors. $\mathbf{F}_i^b$ is the body (e.g., gravitational) force acting on particle $i$.  In the preceding expressions, $W$ is the "smoothed" Dirac delta function with compact support $h$, which integrates to one, and in the limit of $h\to 0$ approaches the Dirac delta function. In this work, we use  $W$  in the form of the fourth-order spline function \cite{Morris1997}:
\begin{align}
W(r,h)= \frac{81}{359 \pi h^3} \left\{
\begin{array}{l l r l l}
\left(3-\frac{3|\boldsymbol{r}|}{h}\right)^5-6\left(2-\frac{3|\boldsymbol{r}|}{h}\right)^5+15\left(1-\frac{3|\boldsymbol{r}|}{h}\right)^5& & 0\leq|\boldsymbol{r}|&<\frac{1}{3}&h\\
\left(3-\frac{3|\boldsymbol{r}|}{h}\right)^5-6\left(2-\frac{3|\boldsymbol{r}|}{h}\right)^5& & \frac{1}{3}h\leq|\boldsymbol{r}|&<\frac{2}{3}&h\\
\left(3-\frac{3|\boldsymbol{r}|}{h}\right)^5& & \frac{2}{3}h\leq|\boldsymbol{r}|&<&h\\
0& & |\boldsymbol{r}|&>&h
\end{array} \right.
\end{align}

In Eq.~(\ref{SDPD-momentum}), $\sum_{j=1}^{N} {F}_{ij}^P$ and $\sum_{j=1}^{N} \mathbf{F}^{visc}_{ij}$  terms are obtained using the SPH discretization of the $\nabla P$ and $\mu \nabla^2\mathbf{v}$ terms in the NS equation, respectively. The $\mathbf{F}_{ij}^S$  force is derived from the expression (\ref{visc-force})  of the viscous/dissipative force using the fluctuation-dissipation theorem instead of directly discretizing the $\nabla \cdot \mathbf{s}$ term in Eq.~(\ref{linear-momentum}).  The number density $n_i = \rho_i / m_i$ can be computed by integrating an ordinary differential equation (ODE) obtained from the SPH discretization of the continuity equation (\ref{Eq-Cont}), but it is more common in SDPD to compute density as
\begin{equation}
\label{eq:smoothed_density}
n_i = \sum_{j} W(r_{ij},h).
\end{equation}

In this work, we use the equation of state
\begin{equation}
P = \frac{c^2\rho_0}{7}\left[\left(\frac{\rho}{\rho_0}\right)^7 -1\right],
\label{eq:EOS}
\end{equation}
where $c$ is the speed of sound and $\rho_0$ is the equilibrium density.

\subsection{Pairwise-Force Smoothed Dissipative Particle Hydrodynamics for low-temperature multicomponent flows}

To impose the boundary condition (\ref{Young-Laplace}), we add the pairwise interaction forces
\begin{equation}
{\bf{F}}_{ij}^{int}= {\bf{F}}^{int}(\mathbf{r}_{ij}) = -s_{ij}\phi(r_{ij}) \frac{\mathbf{r}_{ij}}{r_{ij}}
\label{int-force}
\end{equation}
into the momentum equation,
\begin{eqnarray}\label{PFSDPD-momentum}
m_i \frac{D \mathbf{v}_{i}}{Dt}  =
\sum_{j=1}^{N}\left( \mathbf{F}_{ij}^P + \mathbf{F}^{visc}_{ij} +  \mathbf{F}_{ij}^S  + \mathbf{F}^{int}_{ij} \right)+\mathbf{F}_i^b,
\end{eqnarray}
where  $\phi(r_{ij})$ is the so-called shape factor, to be defined later, and
\begin{equation}
s_{ij}
 = \left\{ {\begin{array}{*{20}{c}}
   s_{\alpha\beta}, &  \; \mathbf{r}_i \in \Omega_\alpha \quad \text{and} \quad  \mathbf{r}_j \in \Omega_\beta,  \\
   s_{\alpha\alpha}, &  \; \mathbf{r}_i \in \Omega_\alpha \quad \text{and} \quad  \mathbf{r}_j \in \Omega_\alpha,  \\
   s_{\beta\beta}, &  \; \mathbf{r}_i \in \Omega_\beta \quad \text{and} \quad  \mathbf{r}_j \in \Omega_\beta.  \\
\end{array}} \right.
\label{Eq-SPH_Int}
\end{equation}

The force ${\bf{F}}_{ij}^{int}$  is a molecular-like pairwise interaction force acting
between particle $i$ of the $\alpha$ phase and particle $j$ of the $\beta$  phase.
The shape function $\phi$ is selected such that ${\bf{F}}_{ij}^{int}$ behaves
as a pairwise molecular force, i.e., ${\bf{F}}_{ij}^{int}$ is short-range
repulsive ($F_{\alpha,\beta}(r_{ij} \le r^* ) < 0$,  $r^* < h$) and long-range
attractive ($F_{\alpha,\beta}( r^*<r_{ij}\le h ) > 0$). For computational
efficiency, $\phi$ should be zero (or decay rapidly) for $r_{ij}\ge h$.
A similar approach to impose the boundary conditions (\ref{Young-Laplace}) has been used
in an SPH multiphase flow model \cite{Tartakovsky2005,Tartakovsky2005b,Tartakovsky2008,Gouet-Kaplan2009,Kordilla2013,Tart-PFSPH}.

Various forms of $\phi$ have been proposed in literature,  and  it has been
shown that $\phi$ affects  particle distribution. Here, we use
\begin{equation}
\phi = r_{ij} \left[-Ae^{-\frac{r_{ij}^2}{2{r_a}^2}}
+ e^{-\frac{r_{ij}^2}{2{r_b}^2}} \right],
\label{eq:pair_interaction}
\end{equation}
where $r_a = r_b/2 = \Delta/2$, which we found to result
in a relatively uniform particle distribution
for a given surface tension value.

It has been demonstrated that, in the absence of thermal fluctuations (i.e., 
for  $\mathbf{F}_{ij}^S = 0$), the parameters in ${\bf{F}}_{ij}^{int}$  can be related to the 
``macroscopic'' surface tension $\sigma_0$ as

 \begin{equation}\label{s11}
 s_{\alpha \alpha }   = s_{ \beta \beta} = 10^k s_{ \alpha \beta} =\frac1{2(1-10^{-k})} n^{-2}_{eq}  \frac{\sigma_0}{ \lambda},
\end{equation}
where $k>1$ and
\begin{equation}\label{T-F-Int}
\lambda  = \frac{\pi}{8}   \int \limits_0^\infty z^4
   \phi(z)dz
   =\pi (-A\varepsilon_0^6+ \varepsilon^6).
\end{equation}

The expression (\ref{s11}) is obtained using the Gibbs treatment \cite{rowlinson2002molecular}, where $\sigma$ is related to the total fluid stress as
\begin{equation}
\label{buff1-general}
\sigma(\mathbf{x}) =\int_{-\infty}^{+\infty} [T_\tau (z) - T_n (z)] dz,\quad \mathbf{x}\in\Gamma.
\end{equation}
Here, $T_n(z)=T_{zz}(z)$ and $T_\tau(z)$ are the normal and tangent components of the stress, and the integration is done along the line crossing $\Gamma$ in the normal direction at point $\mathbf{x}$.

The stress $\mathbf{T}$  can be found in terms of the forces acting between SDPD particles according to the Hardy formula \citep{hardy1982formulas}:

\begin{equation}
{\mathbf{T}}(\mathbf{x}) = {\mathbf{T}}_{(c)}(\mathbf{x}) + {\mathbf{T}}_{(int)}(\mathbf{x}),
\end{equation}
where ${\mathbf{T}}_{(c)} (\mathbf{x})$ is the convection stress,
\begin{equation}
\label{convec-stress}
{\mathbf{T}}_{(c)} (\mathbf{x}) = -
 \sum_{j=1}^N m_j (\overline{\mathbf{v}}(\mathbf{x}) - \mathbf{v}_j    )
 \otimes (\overline{\mathbf{v}}(\mathbf{x}) - \mathbf{v}_j    )
 \tilde\psi_\eta(\mathbf{x}-\mathbf{r}_j),
\end{equation}
and ${\mathbf{T}}_{(int)} (\mathbf{x})$ is the interaction stress,
\begin{equation}
\label{hardy-stress}
\mathbf{T}_{(int)}(\mathbf{x}) =
\frac 12 \sum_{i=1}^N\sum_{j=1}^N \mathbf{f}_{ij}\otimes (\mathbf{r}_j-\mathbf{r}_i)\int_0^1\tilde\psi_\eta(\mathbf{x}-s\mathbf{r}_i-(1-s)\mathbf{r}_j) ds,
\end{equation}
where
$\overline{\mathbf{v}}(\mathbf{x}) = \sum_j m_j\mathbf{v}_j  \tilde\psi_\eta(\mathbf{x}-\mathbf{r}_j)
\left( \sum_j m_j \tilde\psi_\eta(\mathbf{x}-\mathbf{r}_j) \right)^{-1}$
is the average velocity and 
$\mathbf{f}_{ij} = \mathbf{F}_{ij}^P + \mathbf{F}^{visc}_{ij} +  \mathbf{F}_{ij}^S  + \mathbf{F}^{int}_{ij} $ is the total force acting between a pair of $i$ and $j$ particles.
The summation here is over all particles, and $\otimes$ denotes a dyadic product of vectors. The weighting function $\tilde \psi(\mathbf{x})$ is a ``smooth" approximation of the Dirac delta function and can be chosen fairly arbitrary. Here, we assume that
$\tilde \psi(\mathbf{x})$ is the product of one-dimensional functions
$\psi_{\eta, l}=\frac{1}{\eta} \psi \left(\mathbf{x} _{(l)}\right)$, where $l=1, 2, 3$
denotes a vector component. The function $\psi_{\eta, l}(\mathbf{x})$ has compact
support $\eta$,  or becomes sufficiently small for $|\mathbf{r}|>\eta$. In our
calculations, we set $\eta=h$.

The surface tension between any two fluids only depends on the properties of the two fluids and
(weakly depends) on the radii of the interface curvature, but not on the fluid velocities.
As shown in Sec. \ref{sec:sigma_curvature}, the modeled surface tension in our model is independent of curvature
smaller than $\frac{1}{2h}$ for a wide range of temperatures.  To derive the relationship
between  ${\bf{F}}_{ij}^{int}$ and $\sigma$, we first assume the two fluids are separated by an
interface with the radii of curvature much larger than $\eta$ (and $h$). The dependence of surface tension 
on large curvatures will be addressed in Section~\ref{sec:sigma_curvature}.
Under the assumption of small curvature, the interface can be locally  treated as flat.
Because the surface tension is independent of flow conditions, without loss of generality, we
consider the system at equilibrium. It should be noted that in the presence of thermal
fluctuations, SDPD particles move, even at equilibrium. However, in the following derivations,
we disregard the mesoscale effects and compute the macroscale surface tension $\sigma_0$, i.e., we assume that $\mathbf{F}^{visc}_{ij} +  \mathbf{F}_{ij}^S$ and $\mathbf{T}_{(c)}$ have
zero net contribution to  $\sigma_0$.
Furthermore,  it
was demonstrated in \cite{Tart-PFSPH} that if the same average particle density is
used to discretize both fluid phases, then the   $\mathbf{F}_{ij}^P$   has exactly
zero contribution to the surface tension. Finally, Eq.~(\ref{buff1-general}) can be replaced with
\begin{equation}
\label{buff1-simpl}
\sigma_0 (\mathbf{x}) =\int_{-\infty}^{+\infty} [\tilde{T}_\tau (z) - \tilde{T}_n (z)] dz,\quad \mathbf{x}\in\Gamma
\end{equation}
where $\tilde{T}_n(z)=\tilde{T}_{zz}(z)$ and $\tilde{T}_\tau(z)$ are the normal and tangent components of the stress
\begin{equation}
\label{pw-stress}
\tilde{\mathbf{T}}_{(int)}(\mathbf{x}) =
\frac 12 \sum_{i=1}^N\sum_{j=1}^N \mathbf{F}_{ij}^{int}\otimes (\mathbf{r}_j-\mathbf{r}_i)\int_0^1\tilde\psi_\eta(\mathbf{x}-s\mathbf{r}_i-(1-s)\mathbf{r}_j) ds.
\end{equation}

The next step in deriving Eq.~(\ref{s11}) and (\ref{T-F-Int}) is to approximate Eq.~(\ref{pw-stress}) with
\begin{equation}
\label{pw-stress-integral}
\tilde{\mathbf{T}}_{(int)}(\mathbf{x}) =
- \frac 12 n^2_{eq} \int_\Omega \int_\Omega g(\mathbf{r}',\mathbf{r}'') \mathbf{F}^{int}(\mathbf{r}' -\mathbf{r}'')\otimes (\mathbf{r}' -\mathbf{r}'') \int_0^1\tilde\psi_\eta(\mathbf{x}-s\mathbf{r}'-(1-s)\mathbf{r}'') ds d\mathbf{r}' d\mathbf{r}'',
\end{equation}
where $g(\mathbf{r}',\mathbf{r}'')$ is the pair distribution function \cite{Allen1989}.
Assuming that
\begin{equation}
g(\mathbf{r}',\mathbf{r}'')
 = \left\{ {\begin{array}{*{20}{c}}
   g_{\alpha\beta}(\left\vert\mathbf{r}'-\mathbf{r}''\right\vert) , &  \; \mathbf{r}' \in \Omega_\alpha \quad \text{and} \quad  \mathbf{r}'' \in \Omega_\beta,  \\
   g_{\alpha\alpha}(\left\vert\mathbf{r}'-\mathbf{r}''\right\vert), &  \; \mathbf{r}' \in \Omega_\alpha \quad \text{and} \quad  \mathbf{r}'' \in \Omega_\alpha,  \\
   g_{\beta\beta}(\left\vert\mathbf{r}'-\mathbf{r}''\right\vert), &  \; \mathbf{r}' \in \Omega_\beta \quad \text{and} \quad  \mathbf{r}'' \in \Omega_\beta,  \\
\end{array}} \right.
\end{equation}
substituting Eq. (\ref{pw-stress-integral}) in Eq. (\ref{buff1-simpl}) and integrating the latter yields
\begin{equation}
\label{sig14}
\sigma_0 =
s_{\alpha \alpha} \frac{\pi}{8} n_{eq}^2 \int_{0}^\infty g_{\alpha \alpha} (r) \phi(r) r^4 dr
+ s_{\beta \beta} \frac{\pi}{8} n_{eq}^2 \int_{0}^\infty g_{\beta \beta}(r) \phi(r) r^4 dr
-2s_{\alpha \beta}\frac{\pi}{8} n_{eq}^2 \int_{0}^\infty g_{\alpha \beta}(r) \phi(r) r^4 dr
.
\end{equation}

Eq.~(\ref{sig14}) is an extension of an expression for the surface tension of  a single-component 
multiphase molecular system (an $\alpha$-liquid in equilibrium with its gas phase)
given in \cite{rowlinson2002molecular}. This expression assumes that molecules are interacting
via a pairwise force $s_{\alpha \alpha} \tilde{\phi} (r)$,   $s_{\beta \beta} = s_{\alpha \beta} = 0$, and
$n_{eq}$ is the particle density of the liquid phase. To make expression (\ref{sig14}) computable,  
$g_{k l}(r)$ ($(k,l) = \alpha, \beta$) must be defined. A simple approximation is $g(r)=1$,
which is equivalent to treating Eq.~(\ref{pw-stress}) as a Riemann sum.  This reduces
Eq.~(\ref{pw-stress-integral}) to an expression obtained by Rayleigh \cite{Rayleigh}
for a surface tension between two fluids made of molecules interacting via the pairwise
force  $s_{k l} \tilde{\phi} (r)$ ($(k,l) = \alpha, \beta$).
Under the assumption $g(r)=1$, Eqs.~(\ref{s11}) and (\ref{T-F-Int})
follow directly from Eq.~(\ref{sig14}). The details of integration in Eq.~(\ref{sig14}) in two spatial dimensions are given in \cite{Tart-PFSPH}, and the extension to the three-dimensional case is straightforward.

\begin{remark}
We emphasize that Eqs.~(\ref{s11}) and (\ref{sig14}) are derived
with the assumption that the interface between $\Omega_\alpha$ and $\Omega_\beta$
is flat, and that these equations should hold for any interface with the
radii of the largest curvature \emph{much larger}
than $h$, the range of the SDPD forces. In the next section, we show that Eqs.~(\ref{s11})
and (\ref{sig14}) accurately describe the relationship between the surface tension
and pairwise forces for macroscopic systems (where thermal fluctuations are
relatively small or absent) characterized by $\sigma_0$. For a mesoscale fluid
system, the interfacial roughness, induced by thermal fluctuations, may affect the surface tension $\sigma$.
In the following section, we quantify the relationship between $\sigma$,  $\sigma_0$,
$k_B T$, and other model parameters.
\end{remark}

\section{Effect of thermal fluctuations: mean field theory analysis
  and numerical error quantification}\label{Validation}
\label{sec:MFT}
In this section, we study  temperature dependence of the surface tension
for the model introduced in Section~\ref{sec:model}. First, we validate the
method by comparing the surface tension from direct simulation 
with the value prescribed by Eq.~(\ref{s11}).
Here, we prescribe ``low'' temperatures (to be rigorously defined in Secton~\ref{sec:scaling_error})
to keep thermally induced fluctuations small and the fluid interface essentially flat.
In this context, the low temperature refers to the macroscopic limit of the SDPD model,
where thermal fluctuations are negligible, rather than the absolute low state of thermal energy.  

Next, we demonstrate that, for higher temperatures,  Eq.~(\ref{s11})
overestimates the surface tension by a value further dependent on $n_{eq}$.

To explore the effect of thermal fluctuations on the surface tension, we construct
an Euler lattice model based on mean field theory and
obtain a universal scaling relationship that accounts for the effect
of interfacial roughness for various thermal fluctuations and model resolutions.
We then propose a semi-analytical formula to relate the surface tension to model parameters
and quantify the numerical error for different temperatures and model resolutions.

\subsection{Low temperature}
\label{sec:low_temp}
We numerically compute the surface tension between two fluids separated by a nearly flat
interface using Eq.~(\ref{buff1-general}). We model a layer of one fluid surrounded by two layers of another fluid with $n_{eq} = 27$, $46.656$, $64$, and $91.125$.
In all of these simulations, the temperature, speed of sound, and surface tension are set to $k_BT = 0.001$, $c = 6.0$, and $\sigma_0 = 2.1$, respectively.  The parameters $s_{\alpha \alpha}$, $s_{\beta \beta}$, and $s_{\alpha \beta}$ are found from Eq.~(\ref{s11}). The temperature is chosen so the fluid interface remains essentially flat. The simulation domain size is
$10\times10\times16$ with fluid $\alpha$ placed between $-3<z<3$
and fluid $\beta$ occupying the rest of the domain. The periodic boundary conditions are used in all directions.

\begin{figure}[!h]
\subfigure[]{
\includegraphics*[scale=0.3]{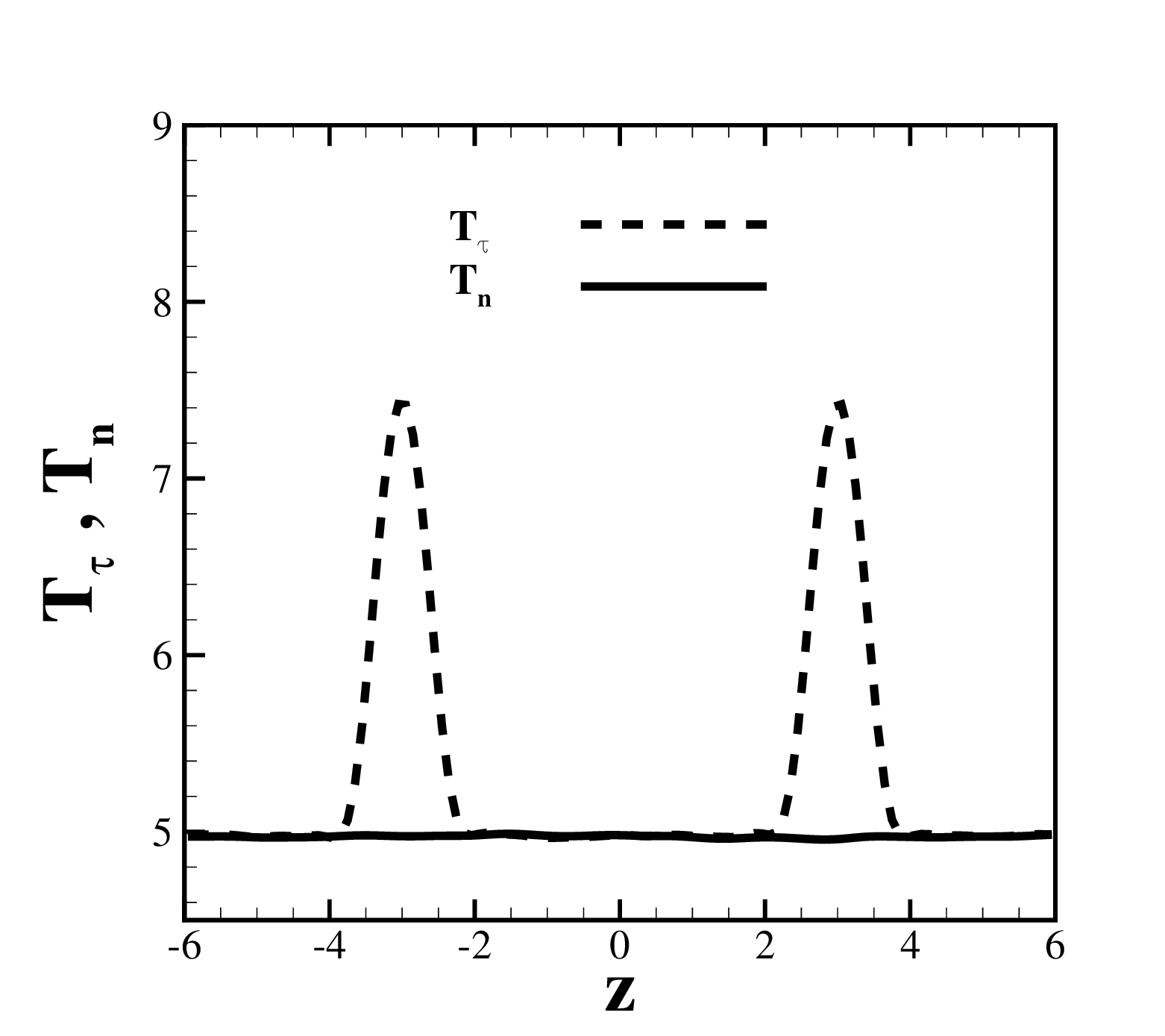}
}
\subfigure[]{
\includegraphics*[scale=0.3]{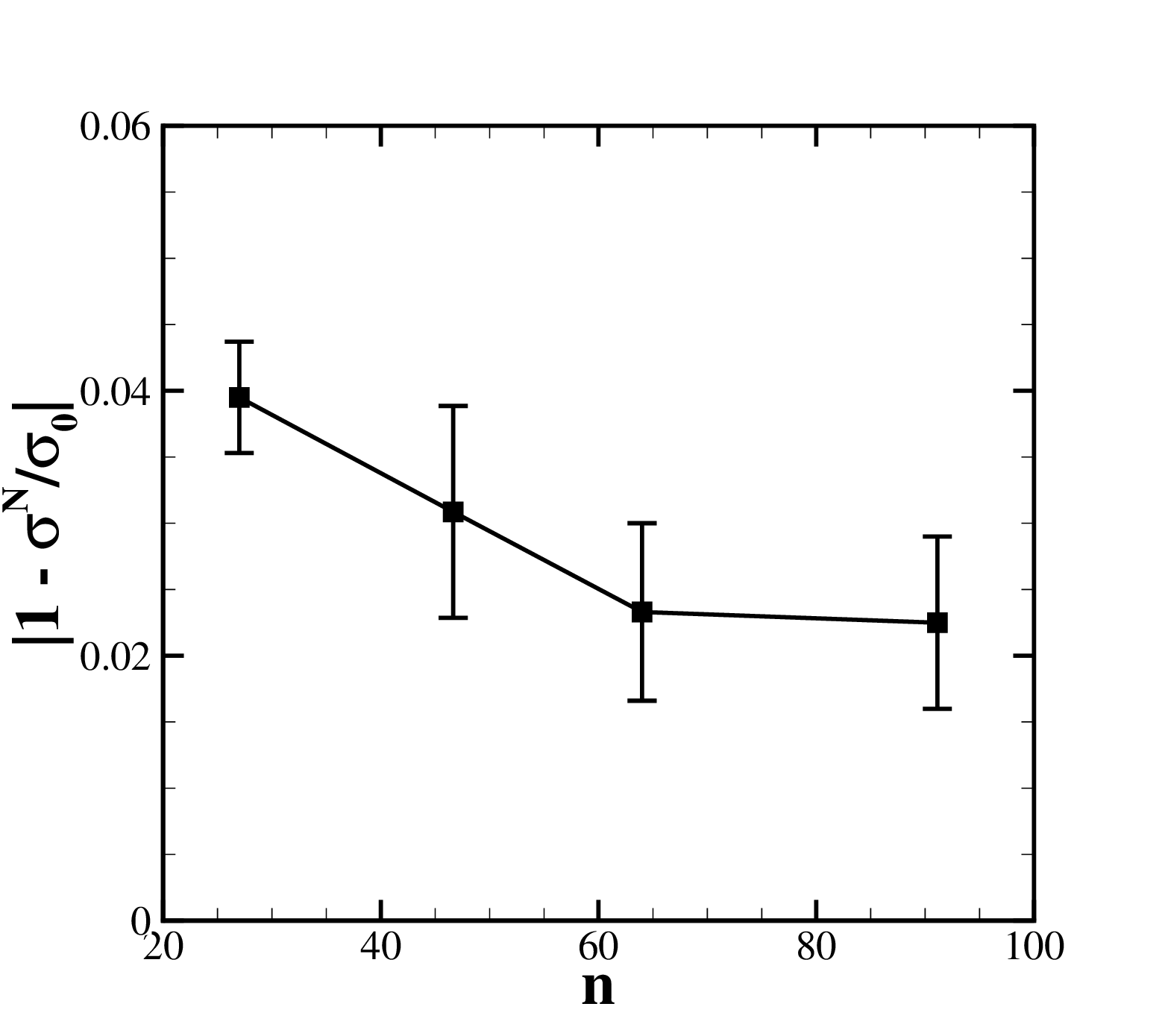}
}
\caption{(a) Normal and tangential hardy stress $T_n$ and $T_{\tau}$ 
  computed across the interface of the two-phase
  fluid layer located at $z = -3.0$ and $z = 3.0$. (b) Numerical error of imposed
 surface tension versus $n_{eq}$ at $k_BT = 0.001$.}
\label{fig:surface_tension_low_temperature}
\end{figure}

Figure~\ref{fig:surface_tension_low_temperature}(a) shows an
example of the normal and tangent  stresses
across the interface for $n_{eq} = 27$.
Similar computations are conducted for other number densities.
For each combination of parameters, $10$ independent simulations are performed,
and $T_{\tau}$ and $T_{n}$ are computed using the simulation data
from the last $10000$ time steps.
Figure~\ref{fig:surface_tension_low_temperature}(b) shows the
relative numerical error, $\varepsilon = |1 - \sigma^N/\sigma_0|$, for different
$n_{eq}$. Here, $\sigma^N$ is the surface tension obtained from direct simulations 
using Eq.~(\ref{buff1-general}).
As $n_{eq}$ increases from $27$
to $64$, $\varepsilon$ decreases from $4.7\%$ to $2.4\%$, respectively. However, $\varepsilon$
does not further decrease as $n_{eq}$ increases from 64 to $91.125$.
One possible reason for this plateau result is the uniform distribution
assumption, $g(r) \equiv 1$, in the derivation of
Eq.~(\ref{s11}), which is not fully
satisfied even for large $n_{eq}$. Nevertheless,
for all considered $n_{eq}$, the error is less
than $5\%$.

\subsection{Effect of thermal fluctuations}\label{thermal-effects}

\begin{figure}[!h]
\includegraphics*[scale=0.3]{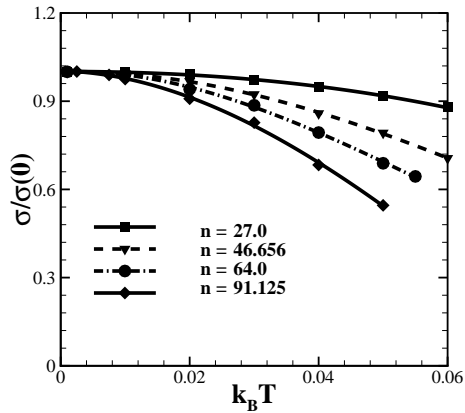}
\caption{Surface tension
between the two-phase flow computed at different temperatures.}
\label{fig:surface_tension_temperature}
\end{figure}

In this section, we model the multiphase system, described in Section \ref{sec:low_temp}, with different temperatures.
Figure~\ref{fig:surface_tension_temperature} shows
the computed surface tension for different $n_{eq}$,
with $k_BT$ between $0$ and $0.06$. As $k_BT$ increases,
the interfacial roughness gets more pronounced (also see Sec. \ref{sec:CW}), and the
surface tension decreases accordingly. Eq.~(\ref{s11}) accurately predicts
the surface tension at the low temperature, but it overestimates the
surface tension at higher temperatures. Moreover, we observe that the
simulated surface tension values, obtained at different $k_BT$, further
depend on $n_{eq}$. Given the same value at the low-temperature
 limit, the surface tension exhibits different temperature-dependent
behaviors for different $n_{eq}$.
For low resolutions (e.g., $n_{eq} = 27$), the
surface tension shows weak dependence on $T$, while for high resolutions, the
surface tension decreases more rapidly as $T$ increases.
To model multiphase flow with thermal fluctuations, we
need to understand the relationship between $\sigma$ and $\sigma_0$, $k_B T$, and $n_{eq}$.

\subsection{Coarse-grained lattice model}
\label{sec:lattice_model}

\begin{figure}[!h]
\includegraphics*[scale=0.5]{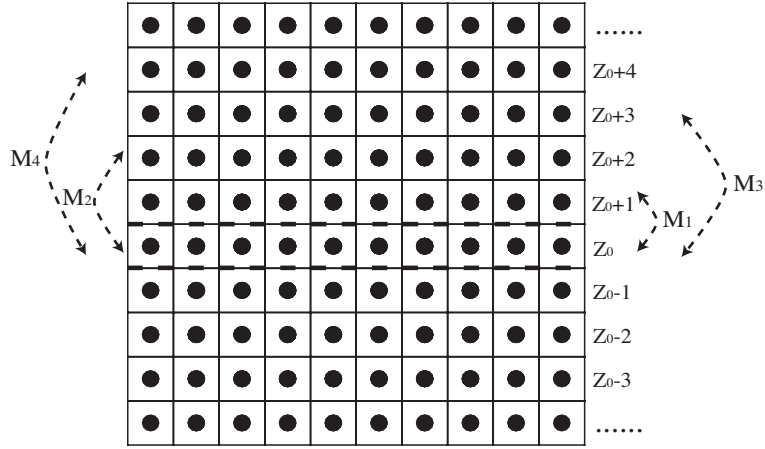}
\caption{A sketch of the coarse-grained lattice model, where the Lagrangian-based
particle model is mapped into a discrete lattice. The lattice layer $z_0$ (e.g.,
the one between the horizontal dashed lines) interacts with the neighboring
layers $z_0\pm1$, $z_0\pm2$, $\ldots$, $z_0\pm L$ with interaction
energy $M_1$, $M_2$, $\ldots$, $M_L$, respectively.}
\label{fig:sketch_lattice_model}
\end{figure}

To quantify the effect of thermal fluctuation on the surface tension,
we coarse grain the system by replacing the present Lagrangian particle system
with a Euler lattice, similar to the work of  
\cite{Widom_JPC_1984, rowlinson2002molecular,dawson1987interfaces}.
Figure~\ref{fig:sketch_lattice_model} shows a sketch of the mapping process. For each
model resolution with $n_{eq} = \Delta^{-3}$ (in three spacial dimensions), we map
the system on the lattice with the lattice size $\Delta$. For each lattice unit, we define
the number density of each lattice $(i,j,k)$ as $n^{L}_{ijk}$. Given a flat
interface, $n^{L} = 1$ within the $\alpha$ fluid region and $0$, otherwise.

For each lattice unit, we
assume the energy can be approximated by a mean field, i.e.,
$u^L = \epsilon n^{L}$, where $\epsilon$ is the potential energy of the lattice
filled with an SDPD particle
\begin{equation}
\epsilon = \int_0^{\infty} 4 \pi u(r) g(r) r^2 dr \approx \int_0^{\infty} 4 \pi u(r) r^2 dr,
\label{eq:energy_unit}
\end{equation}
where $u(r)$ is the potential energy between two fluid particles due to the interaction force, $d u(r)/ dr = s_{\alpha \alpha} \phi(r)$.  Similar to 
Ref. \cite{rowlinson2002molecular},  we
define the activity $\zeta(n^{L}, k_BT)$ for each lattice unit, such that the probability to fill the lattice
with an SDPD particle is
\begin{equation}
n^{L}/\zeta = (1-n^{L})e^{-u^L/k_BT}.
\label{eq:fill_prob}
\end{equation}
Therefore, the activity of each lattice unit is given by
\begin{equation}
\zeta(n^L, k_BT) = \frac{n^{L}}{1 - n^{L}} e^{\epsilon n^{L}/k_BT}.
\end{equation}
Under equilibrium conditions, the equilibrium activity 
$\zeta$ satisfies the equal area rule \cite{rowlinson2002molecular}
\begin{equation}
\int_{n^L_g}^{n^L_l} \ln\left[\frac{\zeta(n^{\prime}, k_BT)}{\zeta}\right]d n^{\prime} = 0, 
  ~~\zeta(n^L_g) = \zeta(n^L_l)
\label{eq:equal_area}
\end{equation}
where $n^L_g$ and $n^L_l$ are the nontrivial solutions ($\neq 0.5$) corresponding to
the coexisting densities of the gas and liquid phases. This gives the equilibrium 
activity $\zeta$ by
\begin{equation}
\zeta = e^{\epsilon/2k_BT}
\label{eq:equ_activity}.
\end{equation}

Next, we consider the inhomogeneous fluid system. Without loss of generality, we assume
the interface is normal to $z$ direction and denote the lattice number density as $n^L(z)$.
As shown in Figure~\ref{fig:sketch_lattice_model}, the lattice unit
at layer $z_0$ interacts with the neighboring
layers $z_0-L, z_0-L+1,\ldots, z_0,\ldots, z_0+L-1, z_0+L$, where $L$
is determined by the cut-off
distance of pairwise force interaction $h$ and lattice unit length $\Delta$ by
\begin{equation}
L = \left \lceil \frac{h}{\Delta} \right \rceil.
\end{equation}

The energy $u^L(z_0)$ of a lattice unit at layer $z_0$ is determined
by the interaction energy with the neighboring layers, which is given by
\begin{equation}
\begin{split}
&u^L(z_0) = \epsilon n^L(z_0) + \sum_{l=1}^{L} M_l \Delta_l^2 n^{L}(z_0) \\
&\Delta_l^2 n^L(z_0)  = n^L(z_0+l) + n^L(z_0-l) - 2n^L(z_0),
\label{eq:mean_field_energy}
\end{split}
\end{equation}
where $\epsilon n^L(z_0)$ represents the energy of the lattice unit under
homogeneous assumption, $M_l$ represents the interaction energy between 
two layers of homogeneous fluid with distance $l \Delta$, and 
$M_l (n^L(z_0+l) - n^L(z_0))$ represents the change of potential 
energy if the number density of the $(z_0+l)th$
layer is changed from homogeneous assumption to $n^L(z_0+l)$. 

$M_l$ is related to the forces acting between SDPD particles, including ${\bf{F}}_{ij}^{int}$, and
can be determined in an iterative manner. As shown
in Figure~\ref{fig:sketch_lattice_model}, we first consider the interaction between
layer $z_0$ and $z_0 + L$. For a single particle in layer $z_0+L$
with distance $z$ to the upper layer $z_0$, the attractive force is
\begin{equation}
\psi(z) = \int_z^{\infty} 2\pi z^{\prime} n_{eq} u(z)dz^{\prime}.
\end{equation}
Therefore, the total interaction force between layers $z_0$
and $z_0+l$ is
\begin{equation}
\theta(z) = \int_{(L-1)\Delta}^{\infty} \psi(z^{\prime}) n_{eq} dz^{\prime},
\label{eq:theta_L}
\end{equation}
and the interaction energy $M_L$ between layers $z_0$
and $z_0 + L$ is 
\begin{equation}
W_L = \int_{(L-1)\Delta}^{\infty} \theta(z^{\prime}) dz^{\prime}, M_L = W_L.
\label{eq:W_L}
\end{equation}
Next, we consider the interaction energy $W_{L-1}$ between the layers
$\left[z_0 -1, z_0\right]$ and $\left[z_0+L-1, z_0+L\right]$. We note
that the interaction energy between layer $z_0$ and $z_0+L$, as well
as layer $z_0-1$ and $z_0+L-1$, is $M_L$. Analysis,
similar to Eqs.~(\ref{eq:theta_L}) and (\ref{eq:W_L}), gives
\begin{equation}
W_{L-1} = \int_{(L-2)\Delta}^{\infty} \theta(z^{\prime}) dz^{\prime}, ~~2 M_L + M_{L-1} = W_{L-1}.
\end{equation}
Repeating the preceding process, $M_l$ can be obtained as
\begin{equation}
\begin{split}
\sum_{k = l}^{L} (k - l +1) M_k = W_l\\
W_l = \int_{(l-1)\Delta}^{\infty} \theta(z^{\prime}) dz^{\prime},
\end{split}
\label{eq:M_l}
\end{equation}
where $M_{l+1},\ldots,M_{L}$ are found iteratively.

Combining  Eqs.~(\ref{eq:energy_unit}), (\ref{eq:equ_activity}), and (\ref{eq:M_l}), we can rewrite Eq. (\ref{eq:fill_prob}) as \cite{Widom_JPC_1984}
\begin{equation}
\begin{split}
&-\sum_{l=1}^L M_l \Delta_l^2 n^L(z) = F^{\prime}\left[n^L(z)\right],\\
&F^{\prime}\left[n^L(z)\right] = -\epsilon \left(\frac{1}{2} - n^L(z)\right)
+ k_BT\ln\left[n^L/(1-n^L)\right].
\label{eq:density_MFT}
\end{split}
\end{equation}
This is the governing equation for the inhomogeneous density
$n^L(z)$ of the coarse-grained lattice model, with asymptotic solutions
$n^L(-\infty) = n^L_g$ and $n^L(\infty) = n^L_l$ satisfying
\begin{equation}
F^{\prime}(n^L_g) = F^{\prime}(n^L_l) = 0.
\end{equation}
By solving Eq.~(\ref{eq:density_MFT}), we can explore the intrinsic relationship
between the Lagrangian particle model and the coarse-grained lattice model, as well as
 quantify the effect of the thermal fluctuations of the surface tension
for different $n_{eq}$, as discussed in Section~\ref{sec:scaling_error}.

\subsection{Effect of thermal fluctuations: scaling and error analysis}
\label{sec:scaling_error}
The numerical solution of Eq.~(\ref{eq:density_MFT}) is complicated by a stiff nonlinear term $k_BT\ln\left[n^L/(1-n^L)
\right]$. To simplify the problem,
we introduce the change of variables
\begin{equation}
n^L(z) = \frac{e^{x(z)}}{1+e^{x(z)}},
\end{equation}
rewrite Eq.~(\ref{eq:density_MFT}) as
\begin{equation}
\begin{split}
&-\sum_{l=1}^L M_l \delta_l^2 {x(z)} = -\epsilon \left(\frac{1}{2} - \frac{e^{x(z)}}
    {1+e^{x(z)}}\right) + k_BT {x(z)}, \\
& \delta_l^2 {x(z)} =  \frac{e^{x(z+l)}}{1+e^{x(z+l)}} + \frac{e^{x(z-l)}}{1+e^{x(z-l)}}
- \frac{2 e^{x(z)}}{1+e^{x(z)}},
\label{eq:density_change_variable}
\end{split}
\end{equation}
and solve it using the Newton-Raphson method on the discrete lattice plane at $z = \ldots, -3/2,
-1/2, 1/2, 3/2, \ldots$. The surface tension of the lattice model
can be determined as
\begin{equation}
\begin{split}
\sigma = \sum_{z=-\infty}^{\infty}\left\{F\left[n^L(z)\right] + \frac{1}{2}
\sum_{l=1}^{L} M_l\left[\Delta_l n^{L}(z)\right]^2\right\} \\
 \Delta_l n^{L}(z) = n^{L}(z+l)-n^{L}(z), ~~ F(n) = \int_{n^L_g}^{n^L_l} F^{\prime}(n) dn.
\end{split}
\end{equation}

To explore the effect of thermal fluctuations on the modeled surface tension, we map the Lagrangian SDPD particles with different
 $n_{eq}$ on a discrete lattice
following the procedure introduced in Section~\ref{sec:lattice_model}. For each $n_{eq}$, we choose $\sigma = 2.1$ at $k_BT = 0.001$ similar to Section~\ref{sec:low_temp}
and solve
Eq.~(\ref{eq:density_MFT}) 
numerically with $\epsilon$  and $M_l$ obtained from Eqs.~(\ref{eq:energy_unit})
and (\ref{eq:M_l}), respectively.

\begin{figure}[!h]
\subfigure[]{
\includegraphics*[scale=0.35]{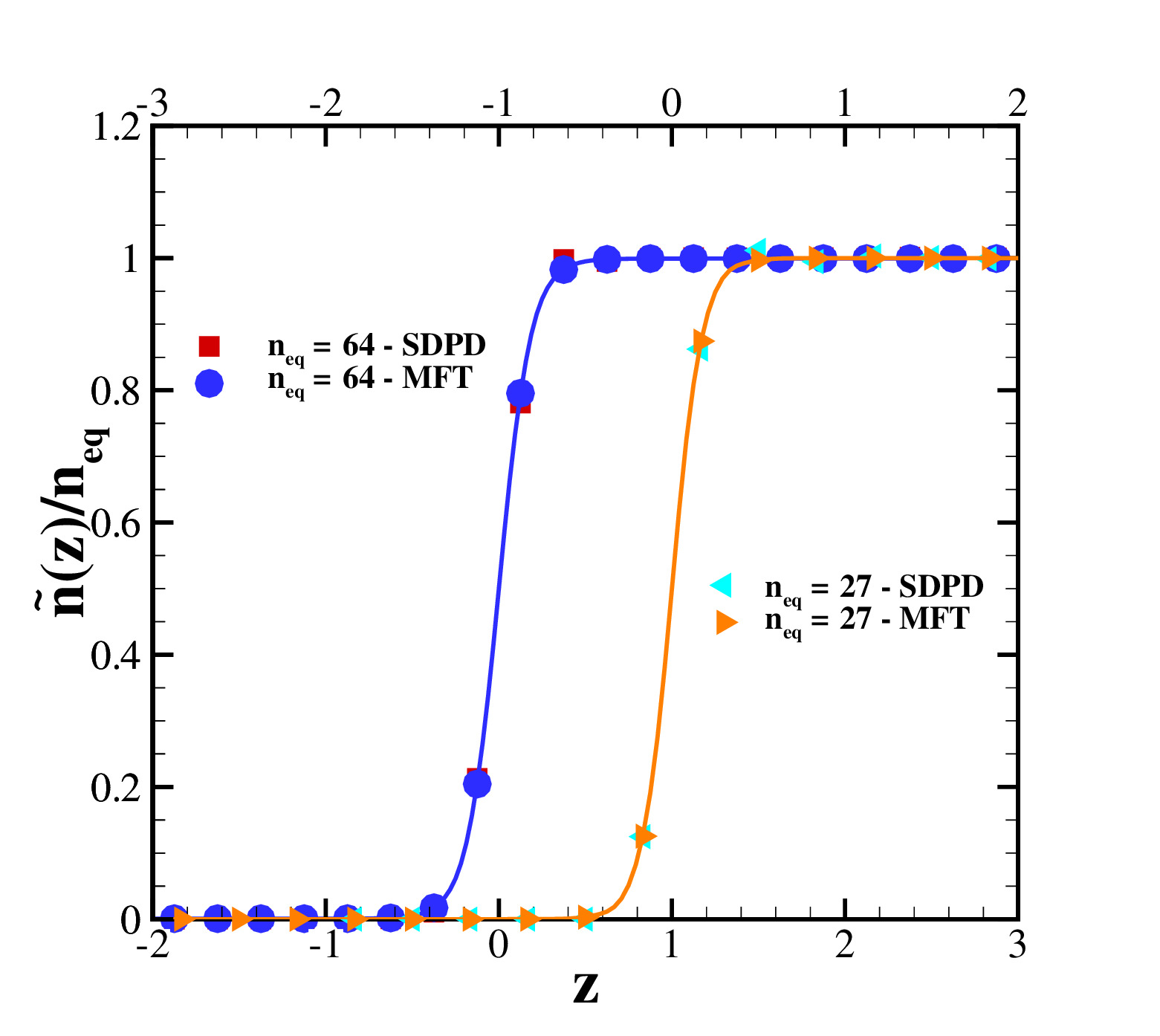}}
\subfigure[]{
\includegraphics*[scale=0.35]{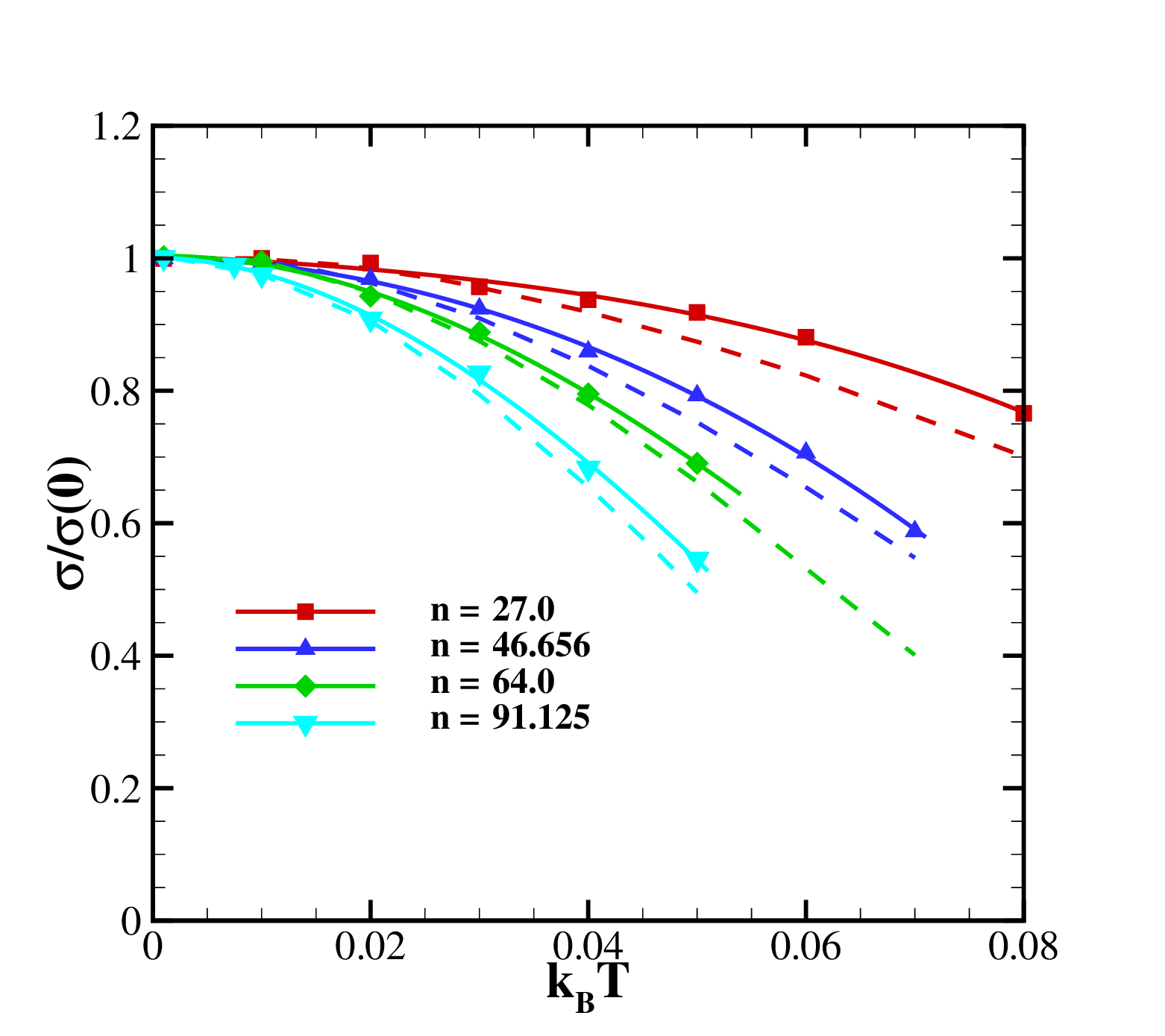}}
\caption{(a) Normalized interfacial density profile $\tilde{n}(z)$
  obtained from both the direct SDPD simulation and the mean field theory
  from Eq. (\ref{eq:density_MFT}) at $k_BT = 0.03$. For the SDPD model, $\tilde{n}(z)$
  is defined as the number density of the SDPD particle with the binning size as
  $\Delta$ along the $z$ direction. (b) Temperature-dependent surface tension obtained
  from the direct simulation and the present lattice model.
$\sigma(0)$ represents the numerical value of the surface tension at $k_BT = 0.001$.}
\label{fig:surface_tension_MFT}
\end{figure}

Figure~\ref{fig:surface_tension_MFT}(a) compares the density profiles obtained from the
the direct simulations and the lattice model with $n_{eq} = 27$ and 64 and $k_BT = 0.03$.
In the lattice model, we numerically solve Eq.~(\ref{eq:density_change_variable}) 
with $\epsilon$  and $M_l$, obtained from Eqs.~(\ref{eq:energy_unit}) and
(\ref{eq:M_l}), respectively. Good agreement is achieved for both $n_{eq} = 27$ and $n_{eq} = 64$.
Figure~\ref{fig:surface_tension_MFT}(b)
shows the surface tensions obtained from the SDPD lattice models
for different $k_BT$ values. It can be seen that the lattice model successfully captures
the  modeled surface tension's dependence on $k_B T$ and $n_{eq}$. 
The difference in $\sigma$ obtained from the  lattice
model and SDPD is mainly due to
the mean field approximation of the energy in Eqs.~(\ref{eq:energy_unit})
and (\ref{eq:mean_field_energy}). As $n_{eq}$ increases, the agreement between
the two models improves. This result not only validates the intrinsic
relationship between the present method and the coarse-grained lattice
model, established through Eqs.~(\ref{eq:energy_unit}), (\ref{eq:M_l}), and
(\ref{eq:mean_field_energy}), but also 
indicates that the lattice model is a convenient tool for
studying the effect of thermal fluctuations on the surface tension in
the present model.

Inspired by the lattice model, we revisit Eq.~(\ref{eq:equal_area}).
It is possible to define the transition temperature
\begin{equation}\label{epsilon}
-\epsilon^L/(k_BT_c)^L = 4,
\end{equation}
above which only the trivial solution exists.  Here, the superscript ``$L$''
 represents the unit lattice length scale $\Delta$.
With $k_BT = k_BT_c^L$, Eq.~(\ref{eq:equal_area}) only has the trivial
solution $n^L_g = n^L_l = 0.5$,
and the surface tension decays to $0$.

We compute $\epsilon$ from Eq.~(\ref{eq:energy_unit})
for different $n_{eq}$ and represent the results in
the lattice model units. 
Because $\epsilon$ and $k_BT$ scale with length unit
$\left[L\right]$ as $\epsilon \sim \left[L\right]^5$ and $k_BT \sim \left[L\right]^2$, where
$\left[L\right]$ represents the length unit of the lattice model,
we have 
\begin{equation}\label{eq50}
-\frac{\epsilon /(\Delta)^5}{k_BT_c / (\Delta)^2} = 4.
\end{equation}
For $n_{eq}  = 27$, $46.656$, $64$,
and $91.125$, the predicted transition temperature $k_BT_c$ is $0.198$, $0.135$, $0.108$, and
$0.087$, respectively.

These results indicate that although the particle model of different
spatial resolution yields the same surface tension when $k_BT$ approaches zero, the
transition temperature of the corresponding lattice model is different, leading to
different temperature-dependent surface tensions for large $k_BT$. Another
way to understand this  is to 
note that the SDPD fluid of different $n_{eq}$ modeled by
Eq. (\ref{s11}) yields the same interaction energy (i.e., $\sum_{l=1}^{L} l M_l$) 
and, therefore, the same surface tension between neighboring layers. However, the
total energy $\epsilon^L$ varies for different $n_{eq}$, leading to a dependence of $\sigma$ on $n_{eq}$.
In particular, smaller $n_{eq}$ yields larger $\epsilon^L$, leading to
a smaller response to interfacial fluctuations. 
Therefore, $\sigma$ decays more slowly for low $n_{eq}$ when $k_BT$ increases.

Based on the preceding analysis, we propose a scaling relationship that
relates the surface tension to the model parameters for different $n_{eq}$, i.e.,
\begin{equation}\label{main-result}
\sigma(k_BT, n_{eq}, \epsilon) = f\left(\frac{k_BT }{n_{eq} \epsilon}\right),
\end{equation}
where $\epsilon$ is determined by Eq.~(\ref{eq:energy_unit}) and the functional form of $f$ depends on the form of $\mathbf{F}^{int}_{ij}$.

Eq.~(\ref{main-result}) is the main theoretical result of this study.
It suggests that the imposed surface tension at $k_BT$, $\sigma(k_BT)$
can be related to the parameters of the rPF-SDPD model, including $k_BT$, $n_{eq}$, and
 surface tension at the ``macroscopic'' interface with zero thermal
fluctuations, $\sigma_0$, through a unified scaling relationship.
For the $\mathbf{F}^{int}_{ij}$ given by Eq.~(\ref{eq:pair_interaction})
and used in this study, we propose an approximate form for $f$ 
\begin{equation}
\sigma^{F}(k_BT, n_{eq}, \epsilon) = \sigma_0\left(1 - b\left(\frac{k_BT}
{ n_{eq} \epsilon  }\right)^2\right)^2,
\label{eq:sigma_fitting}
\end{equation}
where $b = 12.4$ is a fitting parameter.

\begin{figure}[!h]
\includegraphics*[scale=0.4]{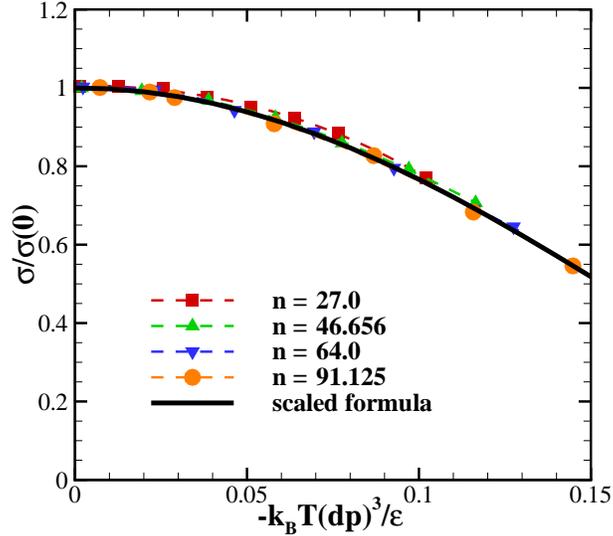}
\caption{Surface tension with respect to $-\frac{k_BT \Delta^3}{\epsilon}$ obtained
  from the direct simulation and Eq. (\ref{eq:sigma_fitting}).
$\sigma(0)$ represents the numerical value of the surface tension at $k_BT = 0.001$.}
\label{fig:surface_tension_scale_formula}
\end{figure}

Figure~\ref{fig:surface_tension_scale_formula} shows the surface tension values $\sigma$ obtained from the rPF-SDPD model and the scaling formula
Eq.~(\ref{eq:sigma_fitting}) as a function of $\frac{k_BT \Delta^3}{\epsilon}$. The simulation
results agree well with the scaling formula for all considered $n_{eq}$, with 
agreement improving with increasing $n_{eq}$.

Table \ref{tab:surface_tension_fit_error} shows the relative differences
$\varepsilon = \left|1 - \sigma^{N}(k_BT)/\sigma^{F}(k_BT)\right|$,
where
$\sigma^{N}$ and $\sigma^{F}$ are the surface tension values obtained
from direct simulation of rPF-SDPD and fitting to scaling relationship
Eq. (\ref{main-result}) through Eq. (\ref{eq:sigma_fitting}), respectively.
For all cases, $\varepsilon$ is less
than $5.7\%$ and decreases with increasing $n_{eq}$.
This shows that Eq. 
(\ref{eq:sigma_fitting}) accurately describes the relationship between the surface tension and $\mathbf{F}^{int}_{ij}$, $n_{eq}$, and $k_B T$.
In the next section, we demonstrate the accuracy and
capabilities of rPF-SDPD by
applying it to several mesoscale multicomponent systems.

\begin{table}[!h]
	\centering
		\begin{tabular} {C{4.5em}|C{7em}|C{7em}|C{7em}|C{7em}|C{7em}|C{7em}|C{4.5em}}
		\hline\hline
		$n$ & $k_BT=0.001$ & $k_BT=0.01$ & $k_BT=0.02$ & $k_BT=0.03$
    & $k_BT=0.04$ & $k_BT=0.05$ & $\left<\varepsilon\right>$\\
    \hline
    $27.0$ & $3.80\%$ & $4.23\%$ & $4.71\%$ & $4.57\%$ & $4.96\%$ & $5.64\%$ & $4.88\%$ \\
    $46.656$ & $3.02\%$ & $3.19\%$ & $3.57\%$ & $3.68\%$ & $3.40\%$ & $4.67\%$ & $3.61\%$ \\
    $64.0$ & $2.15\%$ & $2.75\%$ & $1.45\%$ & $2.44\%$ & $1.63\%$ & $1.47\%$ & $2.17\%$ \\
    $91.125$ & $2.09\%$ & $1.98\%$ & $1.47\%$ & $0.80\%$ & $2.59\%$ & $3.17\%$ & $1.49\%$ \\
    \hline\hline
	\end{tabular}
\caption{Relative error $\varepsilon = \left|1 - 
  \sigma^{N}(k_BT)/\sigma^{F}(k_BT)\right|$ between the
  direct simulated surface tension $\sigma^{N}(k_BT)$ and
  the prediction $\sigma^{F}(k_BT)$ from the scaling relation Eq. (\ref{main-result})
and Eq.~(\ref{eq:sigma_fitting}).
$\left<\varepsilon\right>$ represents the average relative error
for the different $k_BT$ presented here.}
\label{tab:surface_tension_fit_error}
\end{table}

\begin{remark}
In the lattice model, we use the transition temperature
$T_c$ to construct a scaling relationship, but we do not simulate the regime of large thermal fluctuations near $T_c$. In practice, we
note that  for $-\frac{k_BT \Delta^3}{\epsilon} > 0.18$,
SDPD particles of one fluid phase begin escaping into the fluid region of the other phase, i.e., the interface between the fluids diffuses, which is outside of the present study's scope.  In this work,
we consider mesoscale immiscible multicomponent flow,
i.e., the flow with a fluctuating
but clearly defined interface.
\end{remark}

\begin{remark}
The scaling expression (\ref{eq:sigma_fitting}) provides
an accurate approximation of direct simulation results as demonstrated in Table~\ref{tab:surface_tension_fit_error}. However, it is not necessarily a unique expression.
Other formulas in terms of  $\frac{k_BT \Delta^3}{\epsilon}$ may be obtained.
\end{remark}

\begin{remark}
In the present study, we assume the interface has a radius of curvature
much larger than $h$ and can be locally approximated as flat. For an interface with a smaller
radii of curvature, the surface
tension may also depend on the local curvature. We
will address this issue in the next section.
\end{remark}

\section{Numerical examples}\label{Examples}
\label{sec:Numeric}
Here, we study the accuracy of the rPF-SDPD model. First, we show that rPF-SDPD yields consistent thermodynamic properties 
for bulk flow with thermal
fluctuation. Next, we simulate a droplet of one fluid surrounded by another fluid and 
quantify the curvature dependence of the modeled surface tension. 
Finally, we study the dynamics of bubble coalescence
and fluctuations of the fluid interface with and without gravity.

\subsection{Thermodynamic properties}\label{sec:Thermodynamics}
We first demonstrate that the present model accurately captures the
thermodynamic properties of a bulk fluid, i.e., that
in the presence of the pairwise forces $\mathbf{F}_{ij}^{int}$, the probability density function (PDF)
of the $x$, $y$, and $z$ velocity components are in good agreement with
the Maxwell-Boltzmann distribution,
\begin{equation}\label{eq:MB}
f_v (v_k) = \left( \frac{m}{2 \pi k_B T } \right)^{\frac32}
\exp\left(-\frac{ m v_k^2}{2 k_B T}  \right)\quad k = x,y,z,
\end{equation}
and the local density fluctuations satisfy
\begin{equation}\label{eq:density_fluctuation}
\frac{\delta n(d)}{\langle n(d) \rangle}= \sqrt{\frac{k_BT}{\langle n(d) \rangle c^2 d^3}},
\end{equation}
 where $\delta n(d)$ and $ \langle n(d) \rangle$ are the standard deviation of density and
 average density within a cubic domain $\Omega_{d}$ with the edge size $d$,  $d^3$ is the volume of $\Omega_{d}$, and $c$ is the speed of sound of the bulk fluid.

\begin{figure}[!h]
\subfigure[]{
\includegraphics*[scale=0.35]{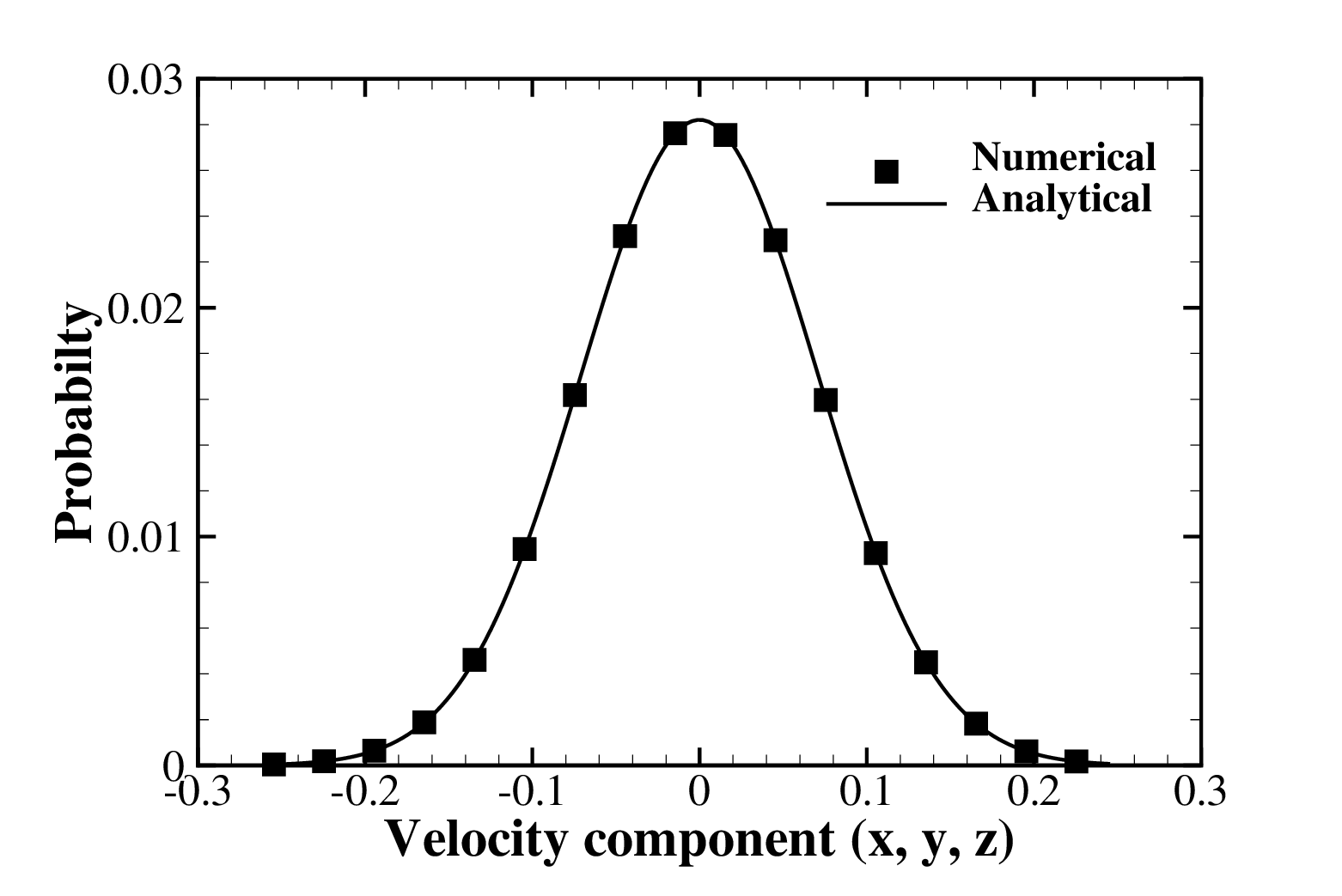}}
\subfigure[]{
\includegraphics*[scale=0.35]{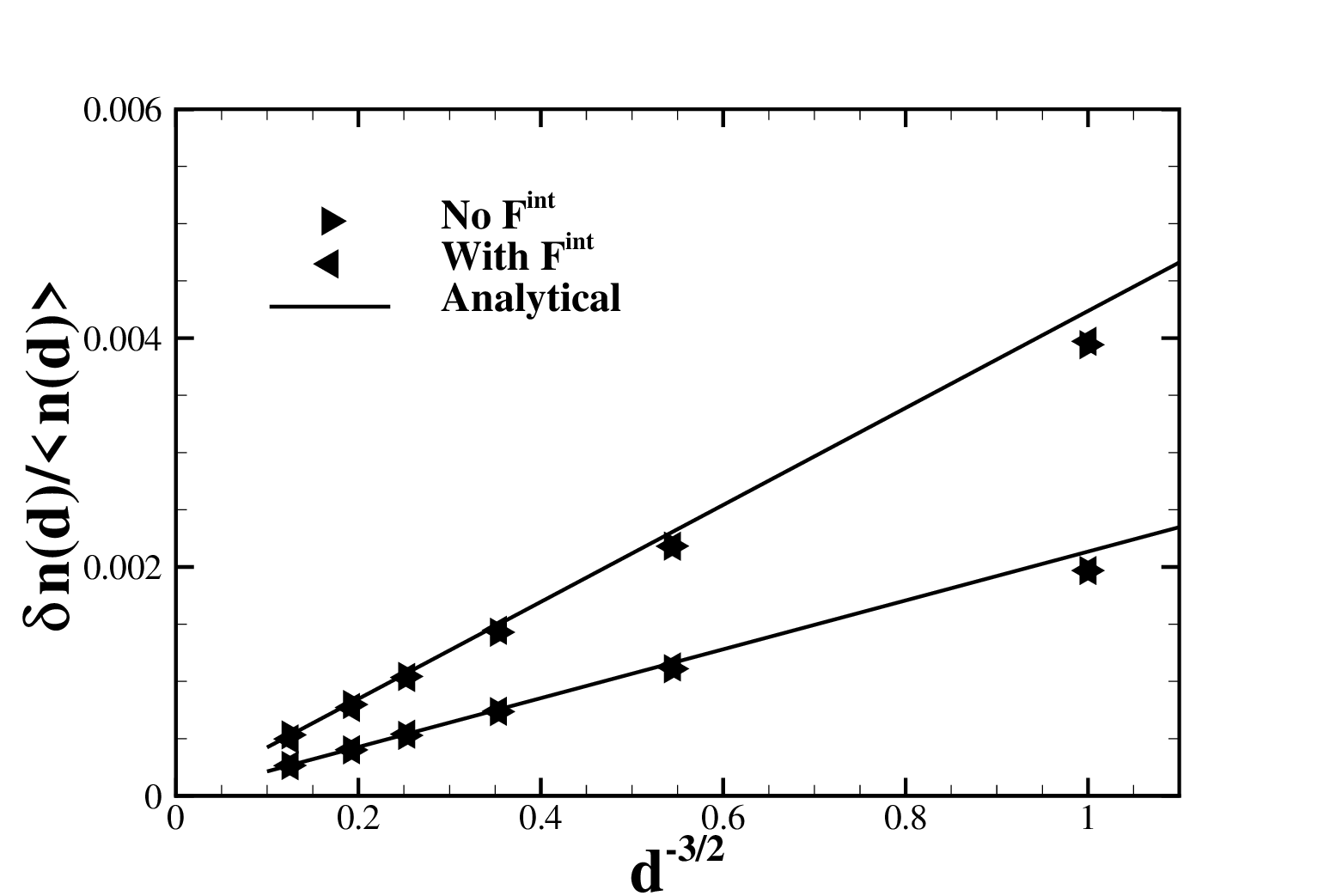}}
\caption{Testing of the mesoscopic method: (a) Probability distribution
  of the velocity components of an individual mesoscopic particle with $k_BT = 0.005$.
    (b) Fluid density fluctuations within different volumes with $k_BT = 0.005$ and
    $k_BT = 0.02$.}
\label{fig:validation}
\end{figure}

We simulate a single-component fluid in a three-dimensional $30\times30\times30 h^3$ 
box in the absence of gravity. The simulations are initialized by placing
particles on a Cartesian mesh with the grid size $\Delta = 0.25 h$. The initial
particle velocity  is set to zero, and the fluid density is set to $n_{eq}= 64 h^{-3}$.
The pairwise forces are chosen to be the same as the numerical example presented in
Section~\ref{sec:low_temp}.
At each time step, we compute the PDFs of the velocity components and local density
$n(d)$ as
\begin{equation}
n(d) = \frac{\sum_{i} n_i \mathrm{I}_{i\in\Omega_{d}}}{\sum_{i}
  \mathrm{I}_{i\in\Omega_{d}}},
\label{eq:n_d}
\end{equation}
where $\mathrm{I}_{i\in\Omega_{d}}$ is an indicator function equal to $1$ if
particle $i$ is inside domain $\Omega_{d}$ and $0$ otherwise; and $n_i$, is found from Eq.~(\ref{eq:smoothed_density}).

Simulations with and without $\mathbf{F}^{int}$ are performed.
Figures~\ref{fig:validation}(a) and \ref{fig:validation}(b)
show that $f_v (v_k)$ ($k = x,y,z$) and  $\delta n(d) /\langle n(d) \rangle$,
obtained from the simulations with and without $\mathbf{F}^{int}$,
agree well with the theoretical results given by Eqs.~(\ref{eq:MB}) and
(\ref{eq:density_fluctuation}), respectively.

For $d = h$,
$\delta n(d) /\langle n(d) \rangle$ is about $6\%$ smaller than the
theoretical prediction from Eq.~(\ref{eq:density_fluctuation}).
This difference arises because $n_i$ in Eq.~(\ref{eq:n_d}) is defined as
a smoothed density  with a smoothing
length $h$. Therefore, for a small volume $\Omega_d$ with
$d$ comparable to $h$, the effective volume is a bit larger than $d^3$,
leading to an underestimation of density fluctuations. Nevertheless,
a good agreement is achieved for $d > 1.6h$. These results demonstrate
that rPF-SDPD yields consistent thermodynamic
properties for the nearly incompressible fluids considered in this study.

\subsection{Effect of interfacial curvature: surface tension of a droplet}
\label{sec:sigma_curvature}
In this section, we compute the surface tension of a three-dimensional droplet of fluid
$\alpha$ immersed in fluid $\beta$. Previous studies show that on the molecular scale,
 surface tension depends on the curvature of the 
interface for the radius of curvature comparable with the molecular size (e.g., see Ref. \cite{kashchiev2003determining}.) 
Since the interaction forces in rPF-SDPD are similar to molecular forces, it is natural to expect the surface tension in rPF-SDPD to depend on the curvature for the radius of the curvature comparable with the particle size or $h$.  In the following, we denote the surface tension
of the droplet of radius $R$ as $\tilde{\sigma}(R, k_BT)$ and explore the
dependence of $\tilde{\sigma}$ on $R$ for various temperatures $k_BT$
and number densities $n_{eq}$.

\begin{figure}[!h]
\subfigure[]{
\includegraphics*[scale=0.3]{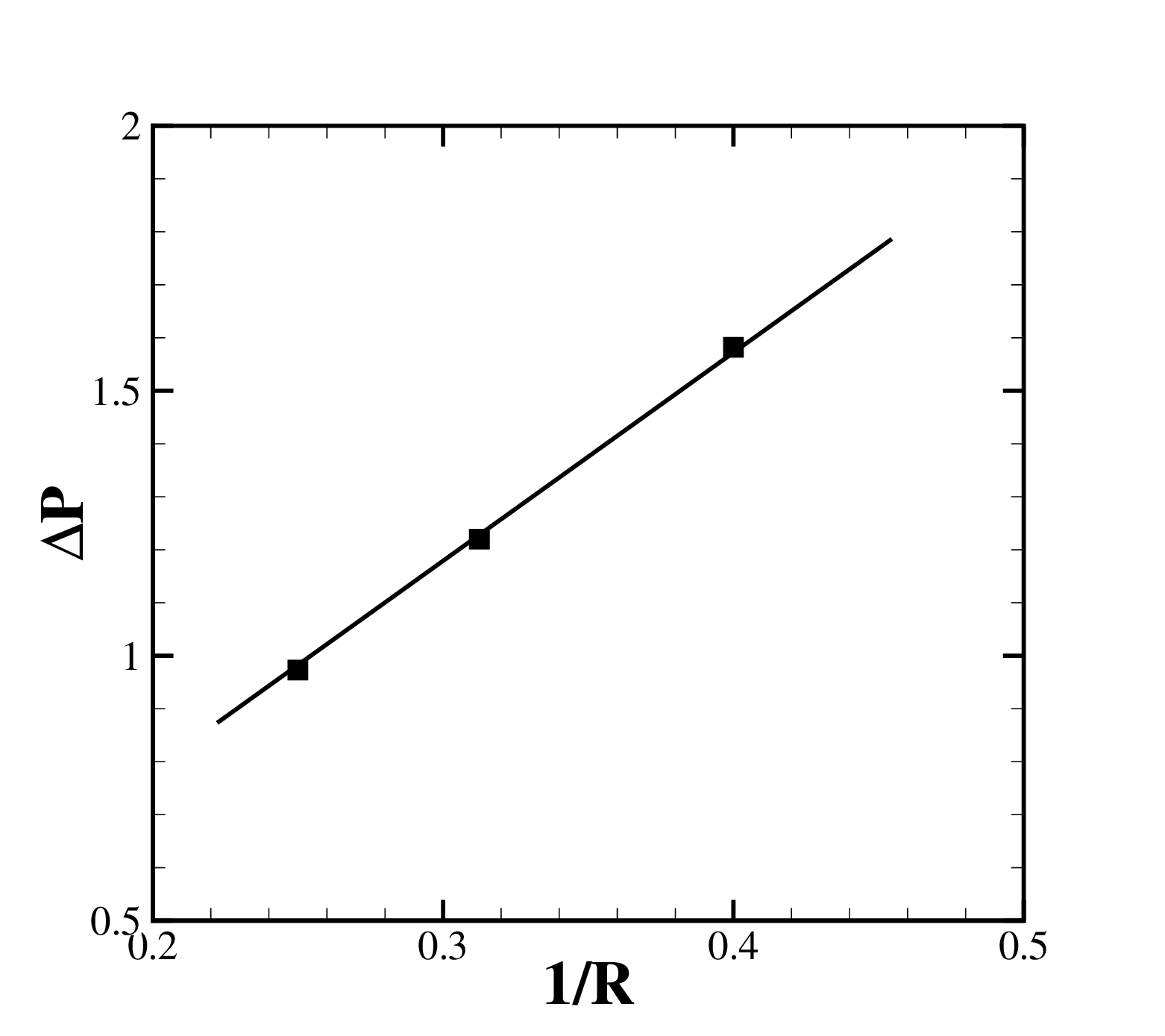}}
\subfigure[]{
\includegraphics*[scale=0.3]{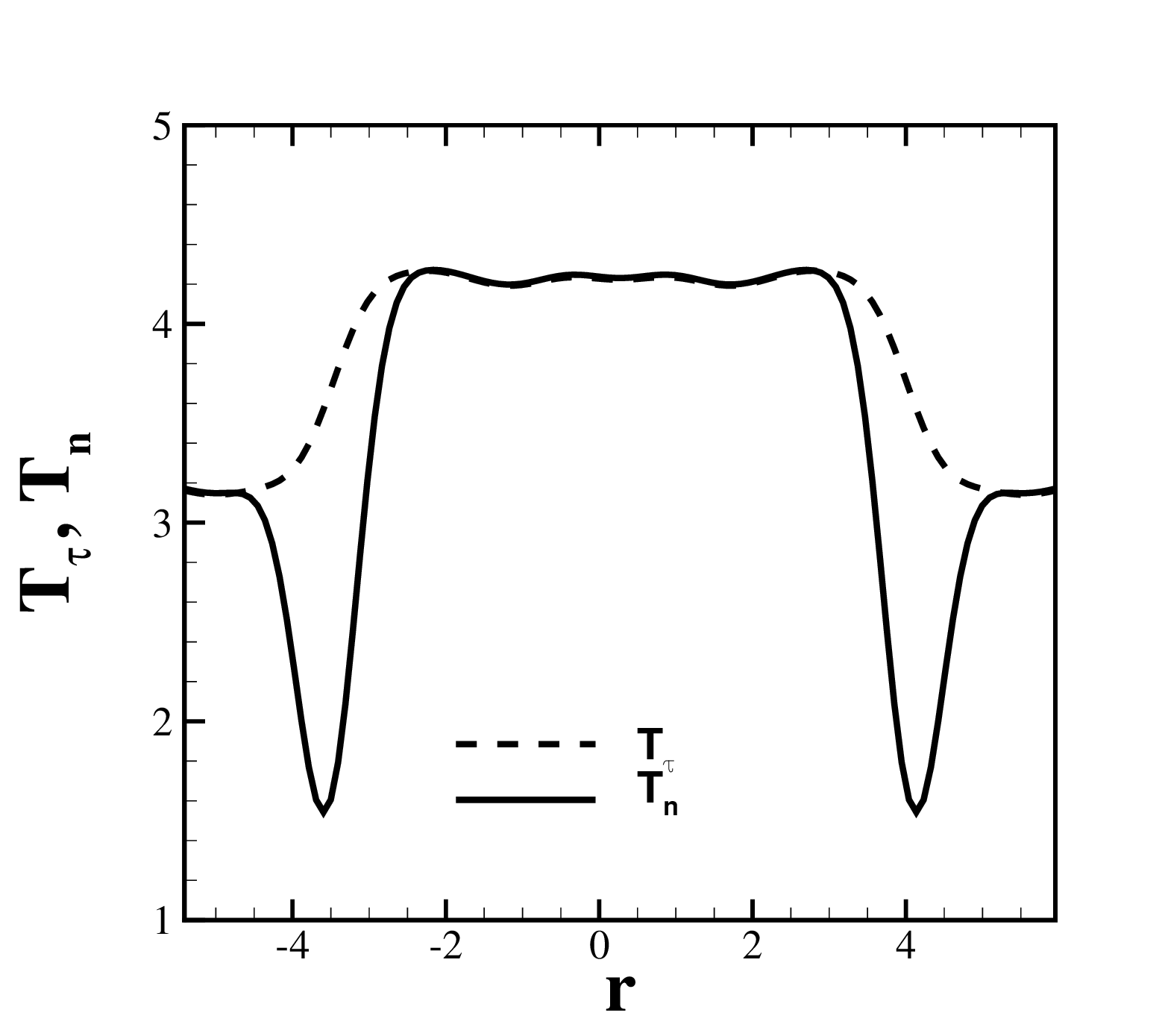}}
\subfigure[]{
\includegraphics*[scale=0.3]{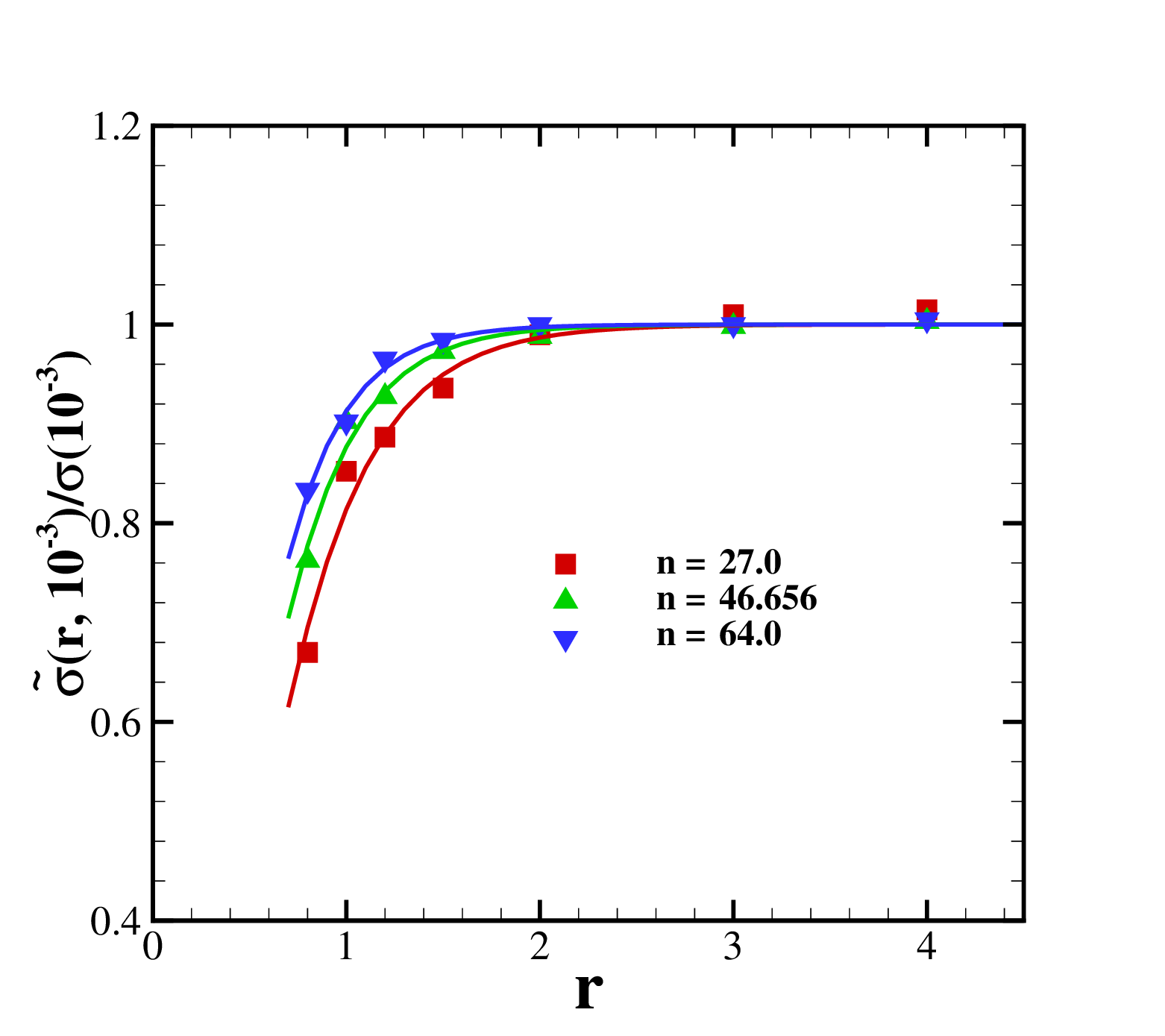}}
\subfigure[]{
\includegraphics*[scale=0.3]{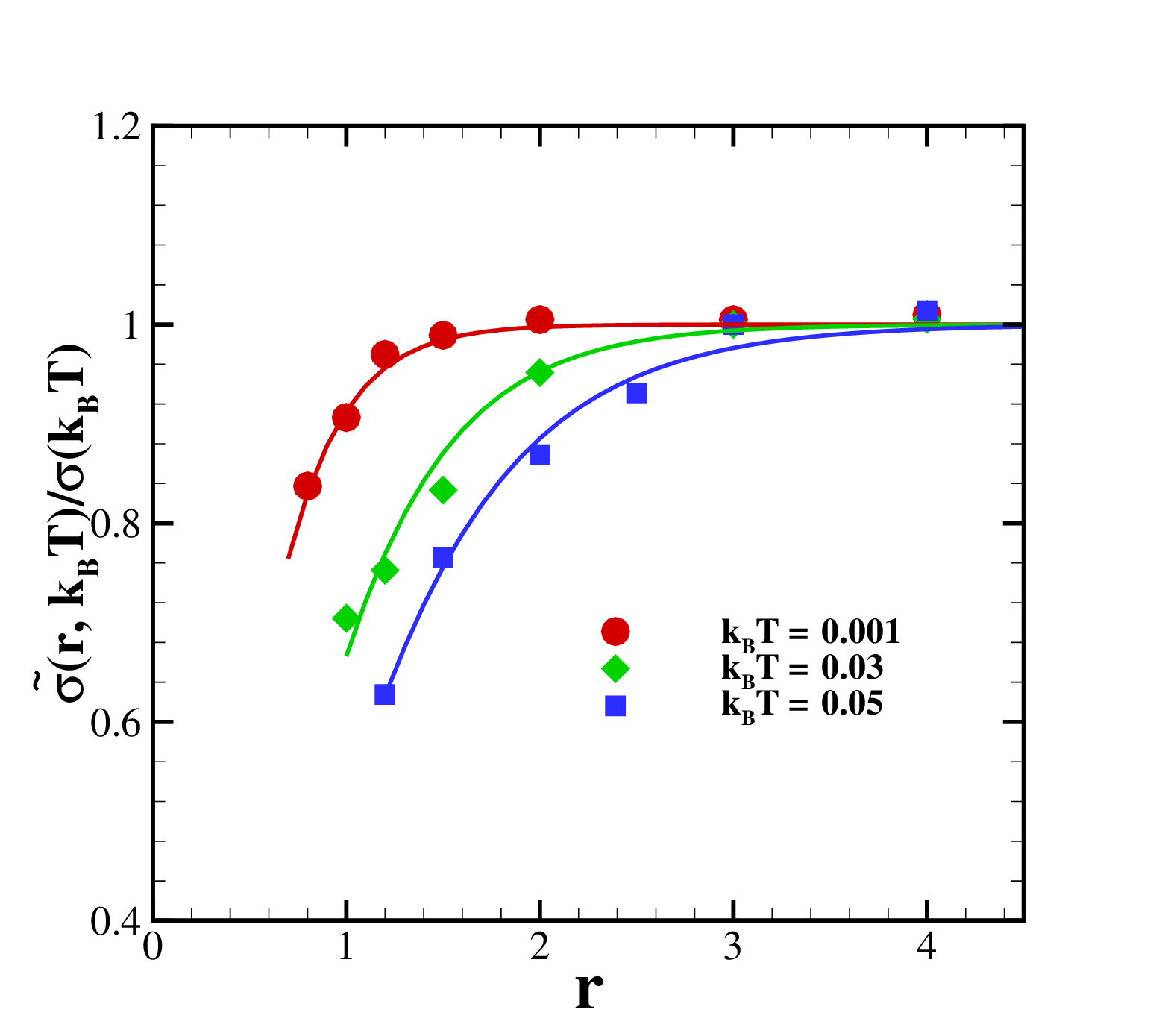}}
\caption{(a) Pressure difference between the inside and outside domain of the droplet
  of radius $R = 2.5$, $3.2$, and $4.0$ with $n_{eq} = 64.0$ and $k_BT = 0.001$.
  The solid line represents the analytical prediction from the 
    Young-Laplace relationship with $\tilde{\sigma}(R) = \sigma = 2.0$.
(b) Normal and tangential hardy stress along the radial direction of the droplet
of radius $R = 4.0$.
(c) Surface tension $\tilde{\sigma}(R)$ computed from droplets
of different radius $R$ with $k_BT = 0.001$. The imposed flat
interface surface tension $\sigma = 2.0$. (d) Surface tension
$\tilde{\sigma}(R, k_BT)$ computed from droplets of different
radius $R$ and $k_BT$ with spatial resolution $n = 64.0$.
The imposed flat interface surface tension is $\sigma(k_BT) =
2.0$, $1.85$, and $1.43$, respectively.
}
\label{fig:droplet_surface_tension}
\end{figure}

First, we  simulate droplets with radii $2.5 h$, $3.2 h$ and $4.0 h$,
$k_BT = 0.001$, and  $n_{eq} = 64.0 h^{-3}$. The computational
domain is $14 \times 14 \times 14 h^{3}$ with the periodic boundary
conditions in the $x$, $y$, and $z$ directions. The droplet is
initially placed in the center of the domain. Pairwise-force parameters
are chosen so the surface tension corresponding to the flat
interface approximation is $\sigma_0 = 2$. We compute the 
surface tension using two approaches,
Eq.~(\ref{buff1-general}) and the Young-Laplace equation
\begin{equation}
\Delta P = \frac{2\tilde{\sigma}(R)}{R},
\end{equation}
where $\Delta P $ is the difference between the pressure inside and outside of the
droplet.
Figures~\ref{fig:droplet_surface_tension}(a) and
\ref{fig:droplet_surface_tension}(b) show $\Delta P$ and the stress components $T_{n}$
and $T_{\tau}$ for the droplet radii greater than $2h$. For all three cases, the
numerical values of $\tilde{\sigma}(R)$ agree well with with the theoretical $\sigma$. These
results show that for large radii $R$, $\tilde{\sigma}(R, k_BT)$
converges to the surface tension between two layers discussed
in Section~\ref{sec:MFT}, i.e., $\tilde{\sigma}(\infty, k_BT) = \sigma(k_BT)$.

Next, we compute the surface tension of the droplets with radii smaller than $4h$ for $n_{eq} = 27$, $46.656$, and $64.0$ at low temperature ($k_BT = 0.001$).
Figure \ref{fig:droplet_surface_tension}(c) depicts the simulation results.
For all $n_{eq}$ and $R<2h$,
$\sigma(R)$ decreases with decreasing $R$. This behavior is similar to that
of the surface tension of  nanoscale droplets, which can
be approximated by \cite{kashchiev2003determining}:
\begin{equation}
\tilde{\sigma}(R) = \sigma \left[ 1 - \exp(-R/R_0) \right],
\end{equation}
where $R_0$ is a radius with the magnitude on the order
of several fluid molecule diameters.
In numerical models, including rPF-SDPD, $\tilde{\sigma}(R)$ is
affected by the resolution when the radius of the curvature is of the same order 
as the spacial model resolution. In the present study, $R_0$ should
be on the order of $h$ or the size of an SDPD particle. In particular,
$\tilde{\sigma}(R)$ converges more quickly to $\sigma$ for larger
spatial resolution $n_{eq}$, and, for all considered $n_{eq}$, $\tilde{\sigma}(R)$
approaches to $\sigma$ for $R = 2.0h$.

We also examine the effect of thermal fluctuations on the curvature dependence of the
surface tension. Figure~\ref{fig:droplet_surface_tension}(d) shows the size-dependent
surface tension of a droplet $\tilde{\sigma}(R, k_BT)$ with
$n_{eq} = 64$ and $k_BT = 0.001$, $0.03$, and $0.05$ for $\sigma(k_BT) = 2.0$, $1.85$, and $1.43$.
The parameters $s_{\alpha \beta}$ in the pairwise force are found from
Eqs.~(\ref{s11}) and (\ref{eq:sigma_fitting}).
As $k_BT$ increases, $\tilde{\sigma}(R, k_BT)$ shows convergence to
$\sigma(k_BT)$ at a slower rate than observed in 
Figure~\ref{fig:droplet_surface_tension}(c). This result is not unexpected, and can be
understood qualitatively as follows: as $k_BT$ increases, the instantaneous
interface exhibits larger deviation from the equilibrium spherical interface
due to larger thermal fluctuations, leading to more pronounced curvature
dependence of the surface tension $\tilde{\sigma}(R, k_BT)$.

Finally, we perform additional simulations with various $n_{eq}$ and observe
 that $\tilde{\sigma}(R, k_BT)$
converged to $\sigma$ for the radii of curvature $R$ satisfying
\begin{equation}
R \ge 2h + 10\left(\frac{k_BT}{\sigma(k_BT)}\right)^{\frac{1}{2}}.
\label{eq:R_criteria}
\end{equation}
To simplify the notation, we  use $\sigma$ to represent $\tilde{\sigma}$ in
the remaining part of the manuscript (if not otherwise specified).

\begin{remark}
We note that Eq.~(\ref{eq:R_criteria}) is an empirical criterion to impose
the prescribed surface tension $\sigma(k_BT)$ for interface with finite
curvature. Theoretical analysis of thermal fluctuation effects 
on $\tilde{\sigma}(R, k_BT)$ is out of scope of the present study.
The implications of Eq.~(\ref{eq:R_criteria}) on numerics will
be discussed in Section~\ref{sec:discussion}.
\end{remark}

\subsection{Bubble coalescence}
\begin{figure}[!h]
\subfigure[]{
\includegraphics*[scale=0.3]{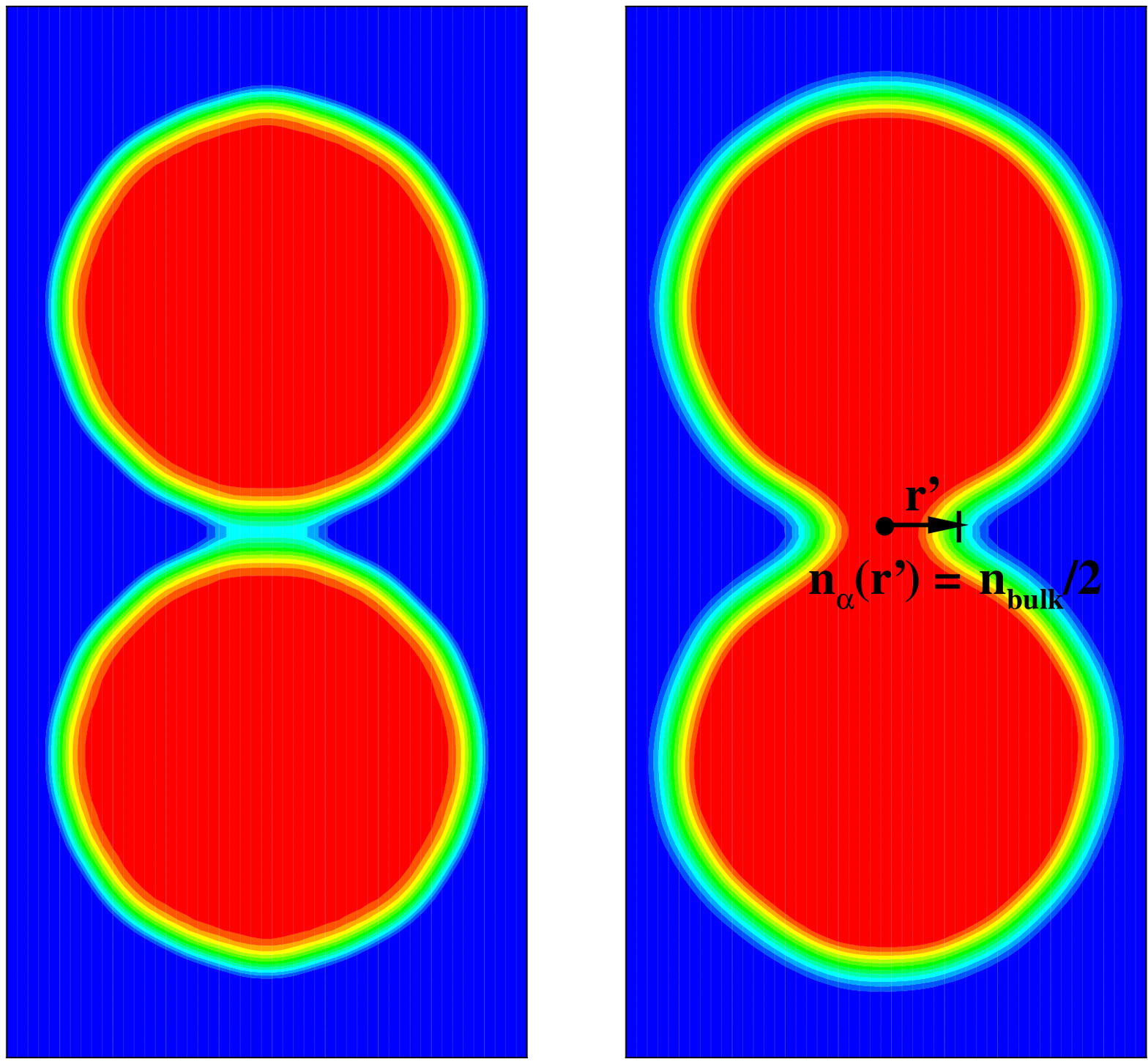}}
\subfigure[]{
\includegraphics*[scale=0.3]{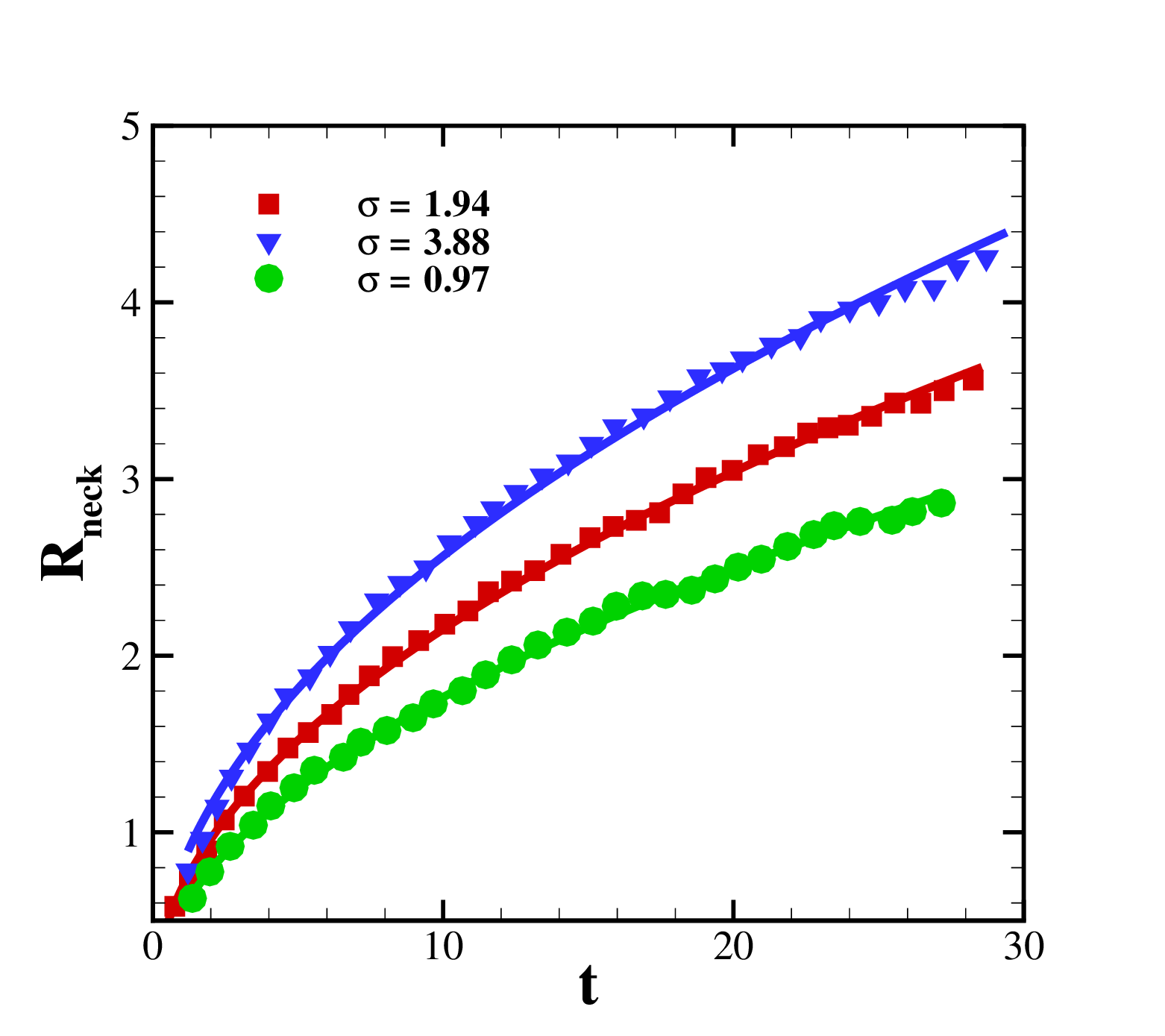}}
\subfigure[]{
\includegraphics*[scale=0.3]{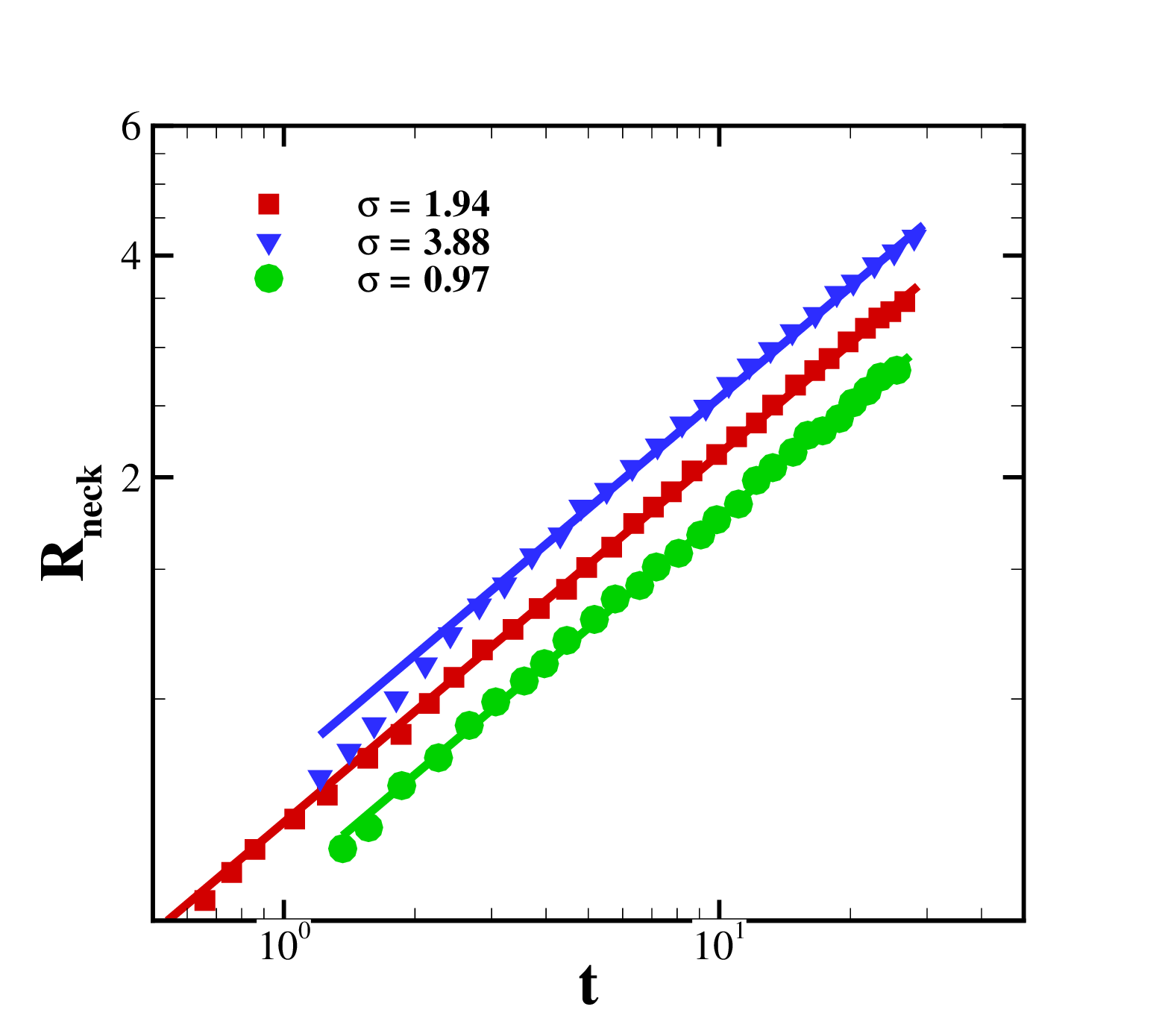}}
\subfigure[]{
\includegraphics*[scale=0.3]{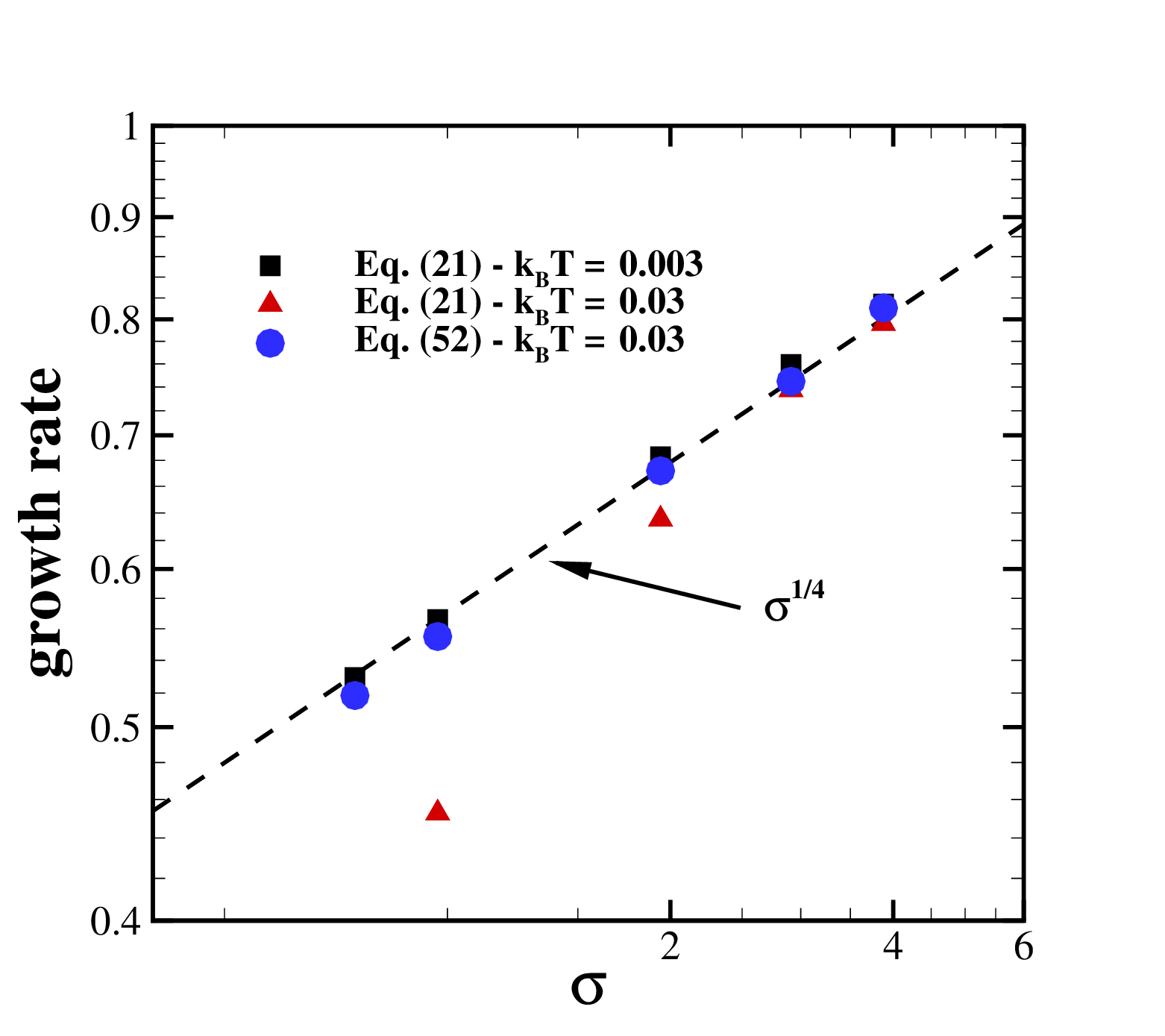}}
\caption{(a) Sketch of the bubble coalescence procedure 
at initial (left) and intermediate (right) stage. For
a specific radial direction $\theta$, $r^{\prime}$
represents the radial distance from center such that 
$n_{\alpha}(r^{\prime},\theta) = n_{bulk}/2$.
(b) The instantaneous neck radius $R_{neck}(t)$ at $k_BT = 0.003$.
(c) $R_{neck}(t)$ plotted in log-log scale.
(d) The growth rate measured with different surface tensions and
$k_BT = 0.003$ and $k_BT = 0.03$.}
\label{fig:bubble_merge}
\end{figure}

Next, we study the dynamic process of bubble coalescence in a
two-phase fluid system similar to Ref.~\cite{PaulSen_Nagel_Nature_2014}.
Two bubbles of radius $R = 4.0 h$ are placed in a simulation domain $24 \times 12 \times 12 h^3$ 
with the centers of the bubbles located at  at $[-4.25 h, 0, 0]$ and
$[4.25 h, 0, 0]$. The periodic boundary condition is used in the simulations.
The spatial resolution $n_{eq}$ and speed of sound $c$ are set to $64.0 h^{-3}$
and $6.0$ for both the bubbles ($\alpha$ fluid) and
surrounding $\beta$ fluid. The mass density $\rho$ is $0.064$ for $\alpha$ 
fluid and $64.0$ for $\beta$ fluid.
The viscosity $\mu$ is $0.005$ for $\alpha$ fluid and $0.5$ for $\beta$ fluid.
As shown in Figure \ref{fig:bubble_merge}(a), we define the instantaneous 
neck radius $R_{neck}$ of the bubble coalescence region by
\begin{equation}
R_{neck}(t) = \frac{1}{2\pi}\int_0^{2\pi}d\theta\int_0^{\infty}dr \delta
\left(n^{\alpha}(\mathbf{r},t) - \frac{n^{\alpha}_{bulk}}{2}\right),
  ~~~~\mathbf{r} = (0, r\cos\theta, r\sin\theta),
\end{equation}
where $n^{\alpha}(\mathbf{r},t)$ is the (time-dependent) smoothed number density of
fluid $\alpha$ at point $\mathbf{r}$ and $n^{\alpha}_{bulk}$ is the bulk number
density of $\alpha$ fluid.  

Under such conditions, the rate of bubble coalescence is controlled by the
surface tension $\sigma$ between $\alpha$ and $\beta$ fluids~
\cite{PaulSen_Nagel_Nature_2014}:
\begin{equation}
R_{neck}(t) = D_1(\sigma R/\rho_{out})^{1/4} t^{1/2} = K t^{1/2},
\label{eq:bubble_growth}
\end{equation}
where $D_1$ is a dimensionless pre-factor, $R$ is the radius of the bubble, 
and $K$ is the growth rate. We
simulate the coalescence process with the surface tension in the range $0.74 < \sigma < 3.88$. 
The radius of the largest curvature satisfies Eq.~(\ref{eq:R_criteria})  in all of these cases.

Figure~\ref{fig:bubble_merge}(b-c) shows the instantaneous neck radius $R_{neck}$
for $k_BT = 0.003$. The growth rate $K$ depends linearly
on $\sigma^{1/4}$, which is consistent with Eq.~(\ref{eq:bubble_growth}).
We also simulate the coalescence process with larger thermal
fluctuations corresponding to  $k_BT = 0.03$. In particular, we impose surface tension
following two approaches: the low-temperature limit given by
Eq.~(\ref{s11}) and the thermal fluctuation
scaling in Eq.~(\ref{eq:sigma_fitting}). The resulting growth
rates are shown in Fig.~\ref{fig:bubble_merge}(d). For large surface tensions, both
models yield consistent growth rates. However,
when $-\frac{k_BT \Delta^3}{\epsilon}$ is large, Eq.~(\ref{s11})
underestimates surface tension and results in slower growth rate.
On the
other hand, the scaling relation Eq.~(\ref{eq:sigma_fitting})
yields a  growth rate consistent
with the theoretical result given by Eq.~(\ref{eq:bubble_growth}) for both $k_BT = 0.003$ and $k_BT = 0.03$.

\subsection{Capillary waves}
\label{sec:CW}
Finally, we examine the interfacial capillary waves in
a two-component fluid system. The entire domain is $[-7,7]\times[-7,7]\times[-10,10] h^3$ with fluid
$\alpha$ placed between $-5h<z<5h$ and fluid $\beta$ occupying the rest of the
domain.
The initial particle density is set to $n_{eq} = 64 h^{-3}$, 
and the speed of sound is set to $c = 6.0$.
In the following, we present results at the interface located at $z = 5h$. 
The results for the interface at $z=-5h$ are identical.
Because of thermal fluctuations, the interface between fluids $\alpha$ and $\beta$ deviates from
a flat plane with the instantaneous height $\eta(x,y)$ defined by
\begin{equation}
n^{\alpha}(x, y, \eta(x,y)) = \frac{n^{\alpha}_{bulk}}{2},
\end{equation}
where $n^{\alpha}(x, y, z)$ is the smoothed density of phase $\alpha$
at  point $(x, y, z)$ and $n^{\alpha}_{bulk}$ is the number density
of $\alpha$ fluid in bulk.

For a fluid system in the absence of gravity, the 
capillary wave theory (CWT) \cite{Buff_Lovett_PRL_1965, Evans_ADP_1979}
predicts that the Fourier modes  (a.k.a. the capillary wave spectra) $\hat{\eta}(\mathbf{q})$ of $\eta(x,y)$ are given by
\begin{equation}
\left<\hat{\eta}(\mathbf{q})^2\right> = \frac{k_BT}{\sigma
  \left\vert \mathbf{q}\right\vert^2 L^2},
\label{eq:CWT}
\end{equation}
where $L\times L$ is the lateral interface domain.
With the external gravity field $g$, acting along the $z$ direction, there is an additional
potential energy change due to the interface fluctuations, e.g., the work
of exchanging the mass density of the lower fluid  $\rho_{\alpha}$
to $\rho_{\beta}$. For each $\mathbf{q}$, the contribution to the potential energy
difference $\Delta H_g$ is given by
\begin{equation}
\Delta H_g = \frac{1}{2} \left\vert\hat{\eta}(\mathbf{q})\right\vert^2 (\rho_{\alpha}
    - \rho_{\beta})g L^2.
\label{eq:potential_contribution}
\end{equation}
Therefore, the variance of $\hat{\eta}(\mathbf{q})$ of the fluctuating interface in the presence of gravity $g$ is given by
\begin{equation}
\left<\hat{\eta}(\mathbf{q})^2\right> = \frac{k_BT}{\sigma \left\vert\mathbf{q}\right\vert^2 L^2 + (\rho_{\alpha}-\rho_{\beta}) g L^2}
\label{eq:CWT_gravity}
\end{equation}

\begin{figure}[!h]
\subfigure[]{
\includegraphics*[scale=0.3]{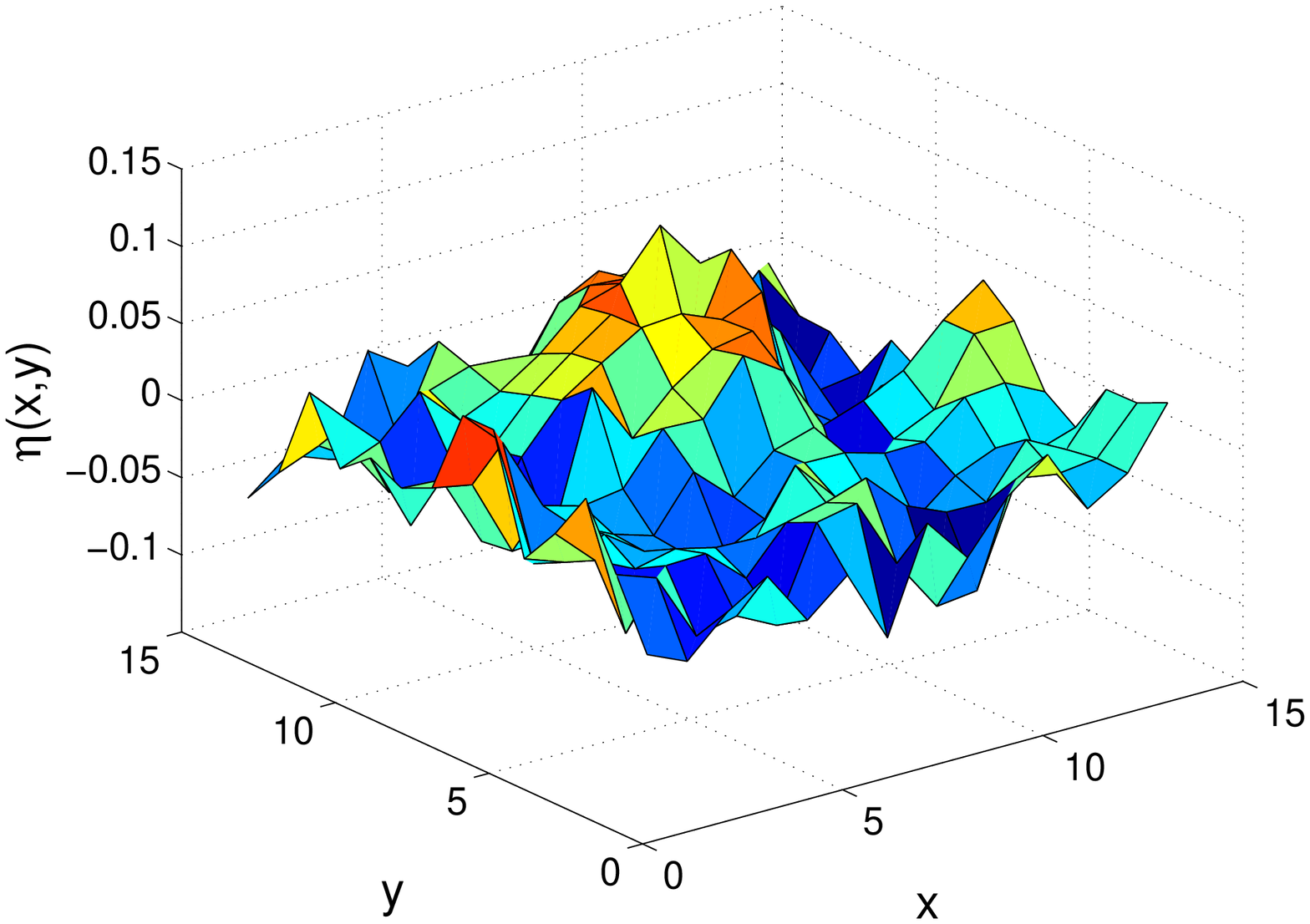}}
\subfigure[]{
\includegraphics*[scale=0.3]{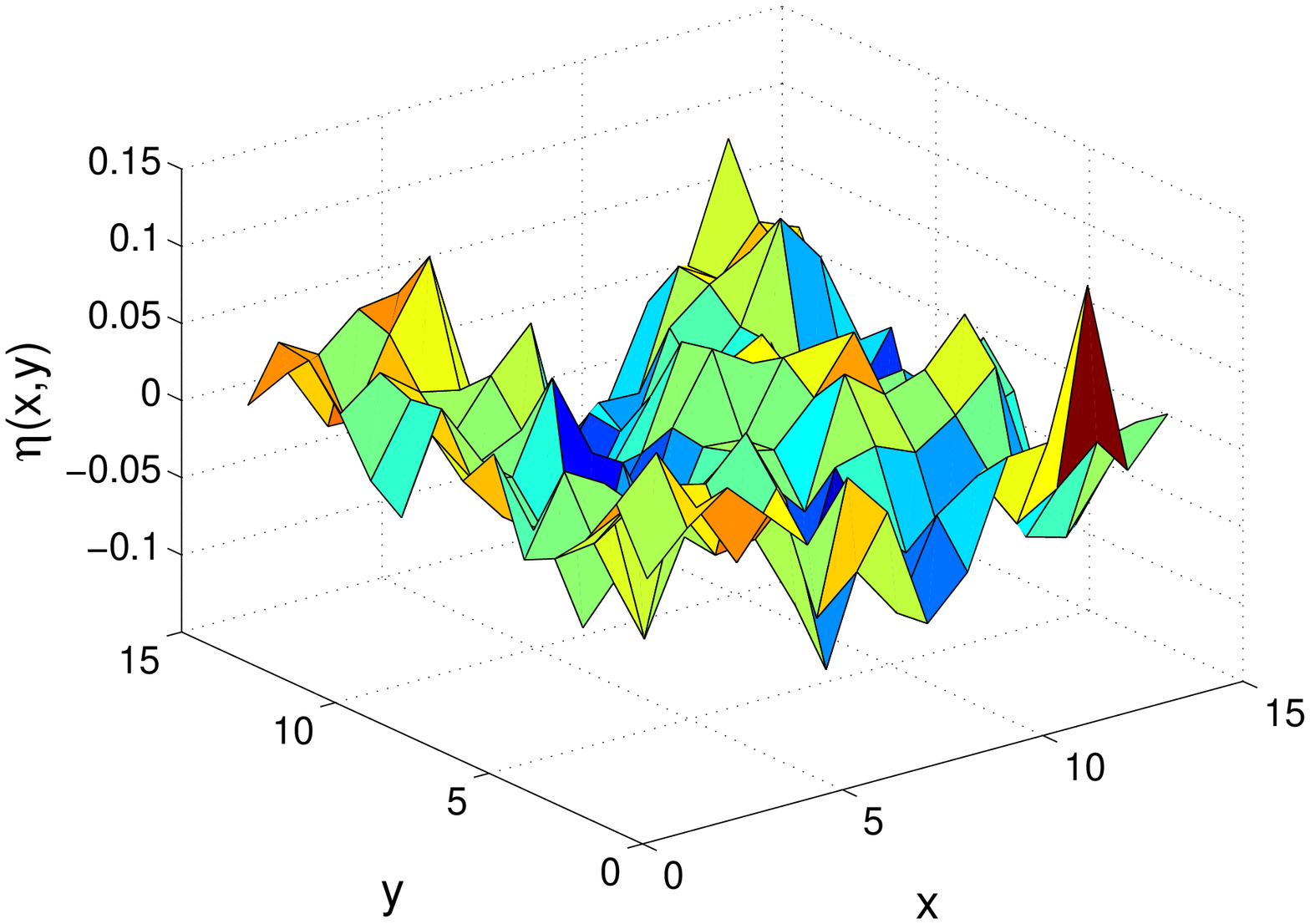}}
\subfigure[]{
\includegraphics*[scale=0.3]{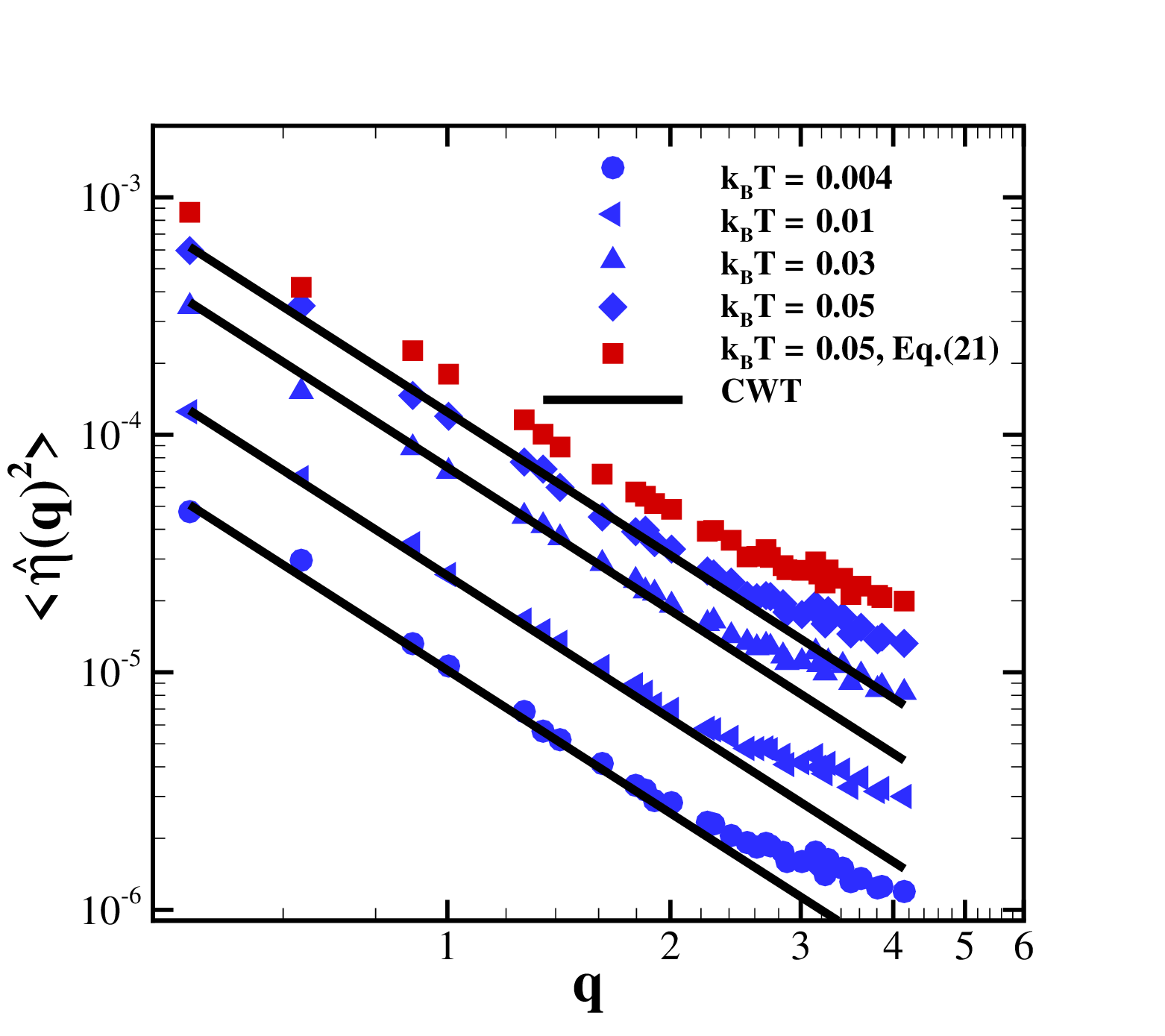}}
\caption{(a) Instantaneous fluid height near the interface at $k_BT = 0.004$.
(b) Instantaneous fluid height near the interface at $k_BT = 0.01$.
(c) The capillary wave spectra measured at different temperatures.
For $k_BT = 0.05$, spectrum with surface tension determined directly from
Eq. (\ref{s11}) is also presented.}
\label{fig:CWT_neutral}
\end{figure}

First, we study the zero-gravity ($g=0$) capillary wave spectra $\hat{\eta}(\mathbf{q})$ with
$\rho_{\alpha} = \rho_{\beta} = 64.0$ and varying
thermal fluctuations. The surface tension $\sigma = 2.0$ is imposed, following
Eq.~(\ref{eq:sigma_fitting}). Figures~\ref{fig:CWT_neutral}(a)
and \ref{fig:CWT_neutral}(b) show  $\eta(x,y)$ with
$k_BT = 0.004$, and $0.01$. As expected, $k_BT = 0.01$ yield larger
interfacial fluctuations. Figure~\ref{fig:CWT_neutral}(c) shows the
spectra $\hat{\eta}(\mathbf{q})$ at $k_BT = 0.004$, $0.01$, $0.03$,
and $0.05$.
For all $k_BT$, $\hat{\eta}(\mathbf{q})$ agrees well with
 Eq.~(\ref{eq:CWT}) for low wave numbers and deviates from
the CWT prediction for $\left\vert\mathbf{q}\right\vert
\ge \frac{2\pi}{5h}$, where $h$ is the support of the kernel $W$.
This discrepancy
is primarily due to the continuum assumption in CWT, where the interfacial
energy is modeled as an increased surface area multiplied by the constant
surface tension. However, for small length scales, local interfacial
energy also depends on the local curvature and interactions
between the SDPD particles (also shown in
Section~\ref{sec:sigma_curvature}).
Therefore, the CWT prediction is not valid for high wave numbers (also see 
\cite{Lei_Schenter_JCP_2015}).
Remarkably, for $k_BT = 0.05$, we also present the spectrum
obtained by imposing surface tension directly from Eq. (\ref{s11});
$\left\vert\hat{\eta}(\mathbf{q})\right\vert^2$ deviates from CWT prediction for 
all $\mathbf{q}$ due to  numerically overestimated interfacial surface tension
at high temperatures.

\begin{figure}[!h]
\subfigure[]{
\includegraphics*[scale=0.3]{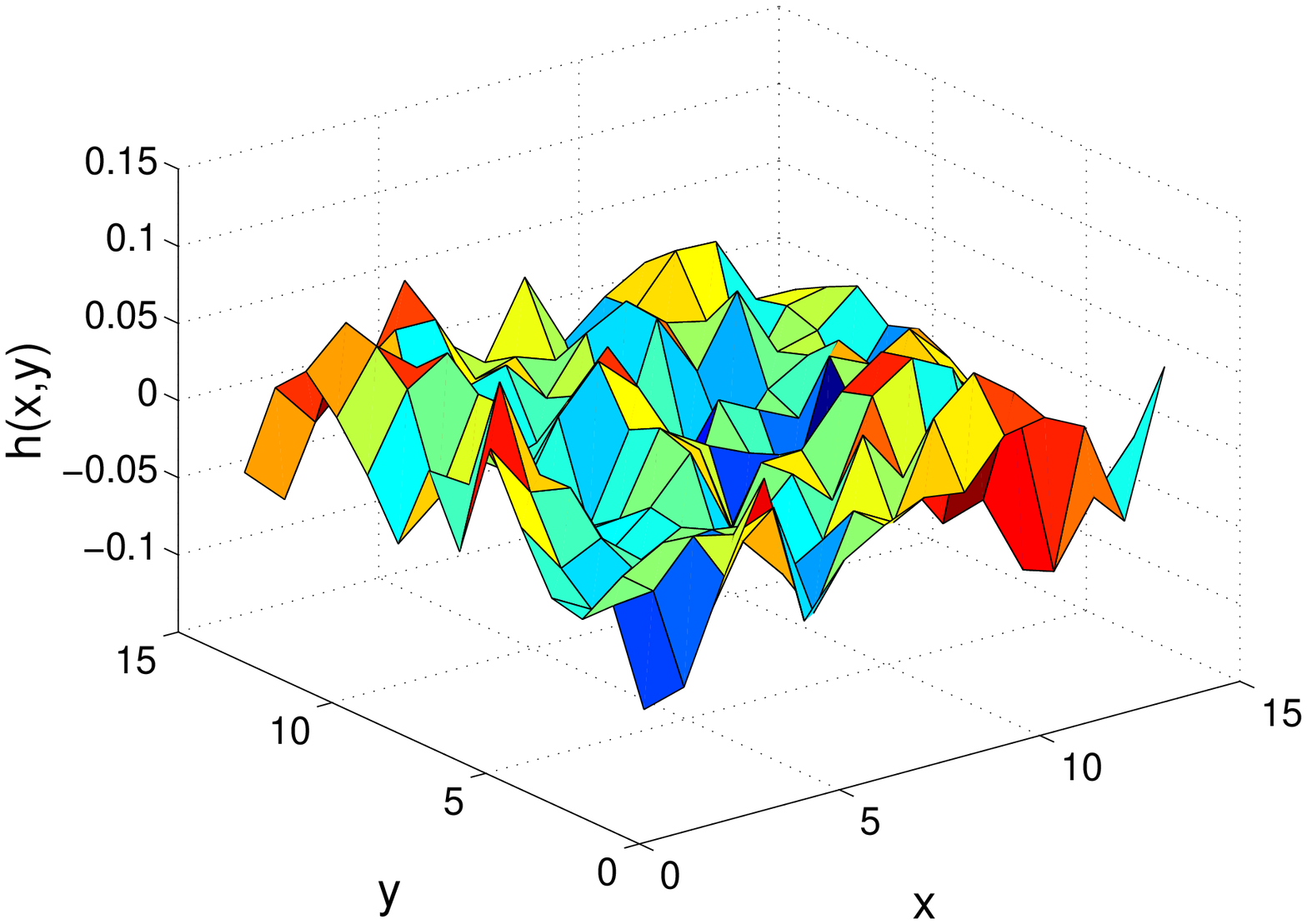}}
\subfigure[]{
\includegraphics*[scale=0.3]{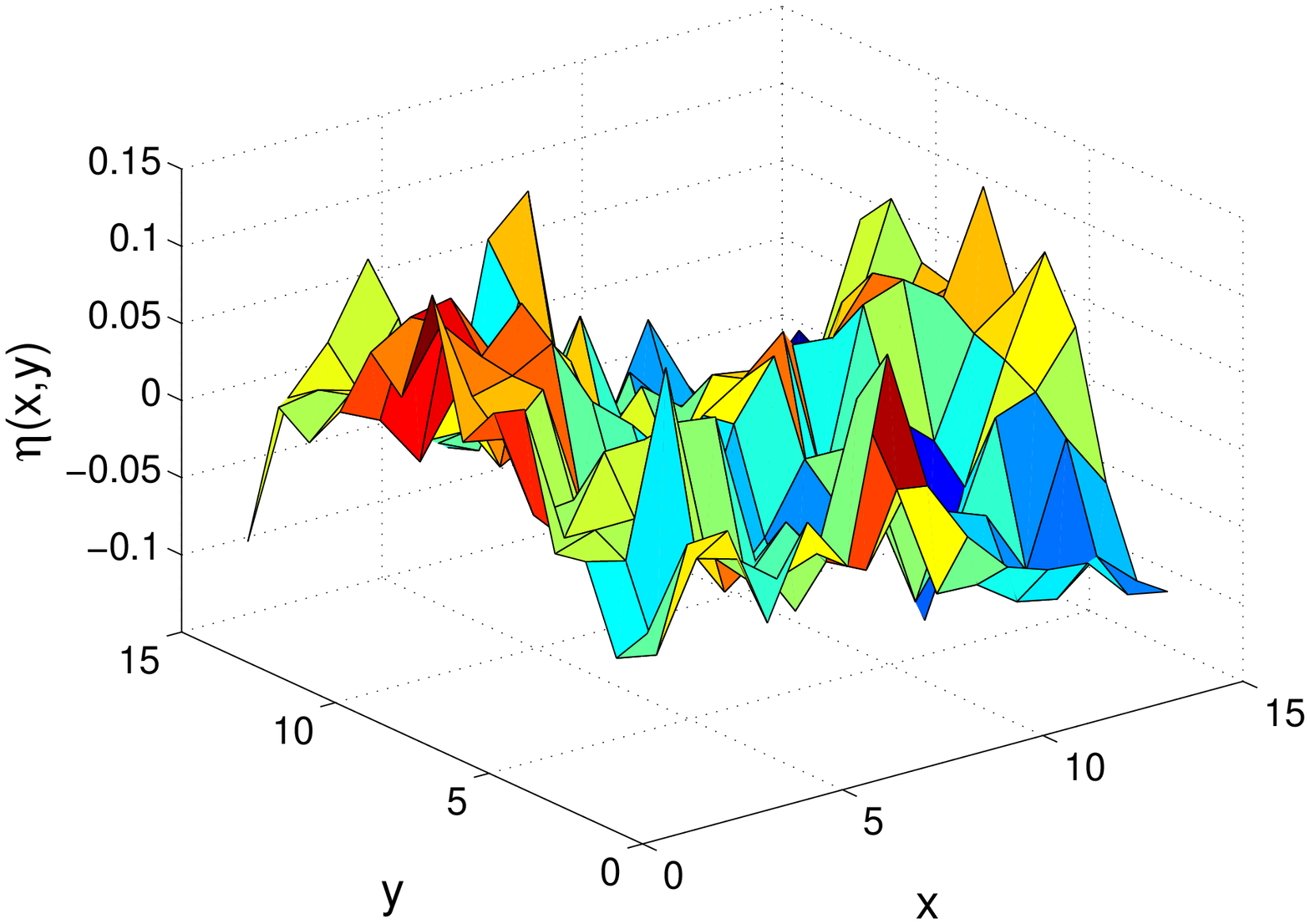}}
\subfigure[]{
\includegraphics*[scale=0.3]{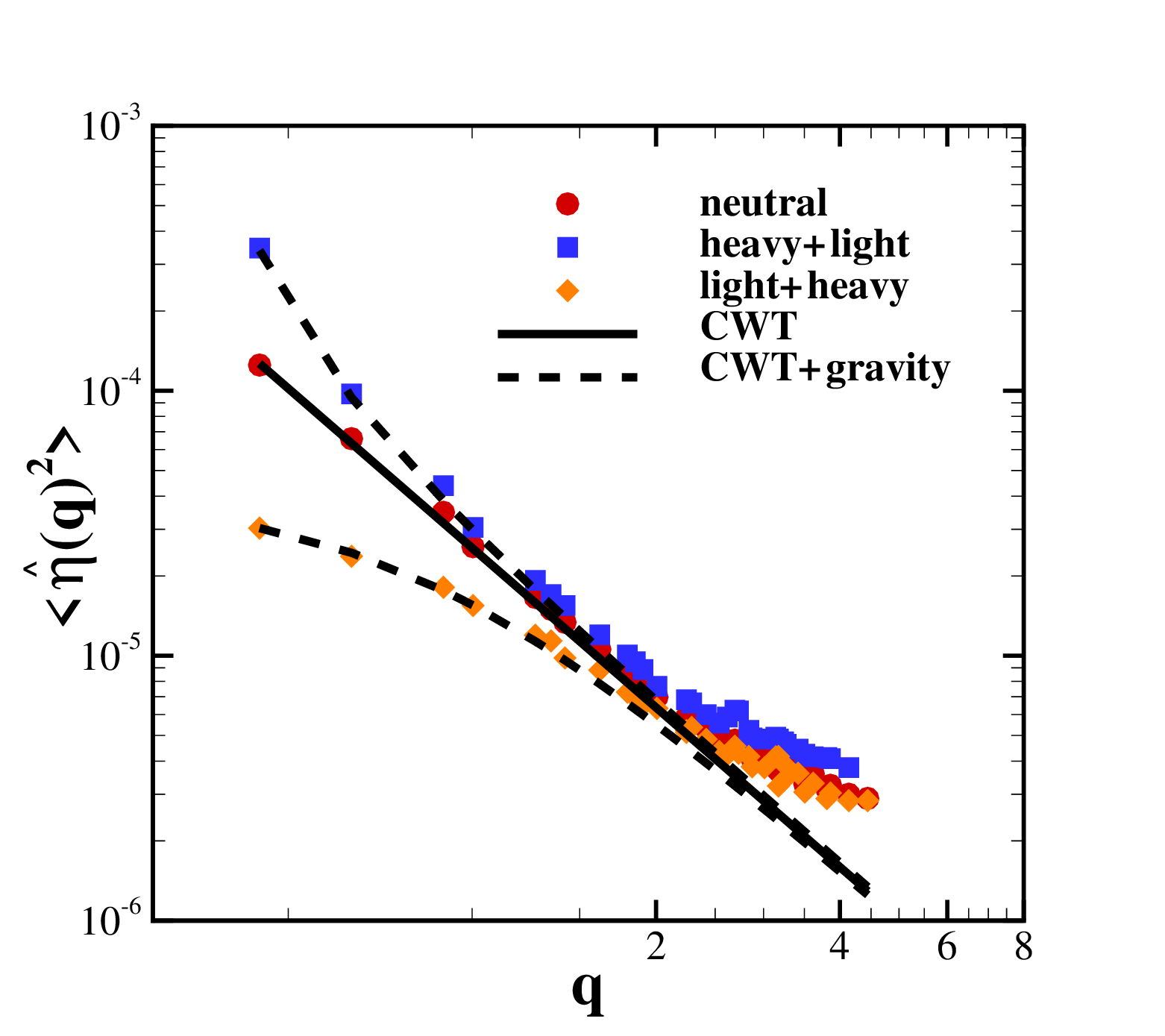}}
\subfigure[]{
\includegraphics*[trim = 5mm 0mm 5mm 20mm,clip,scale=0.3]{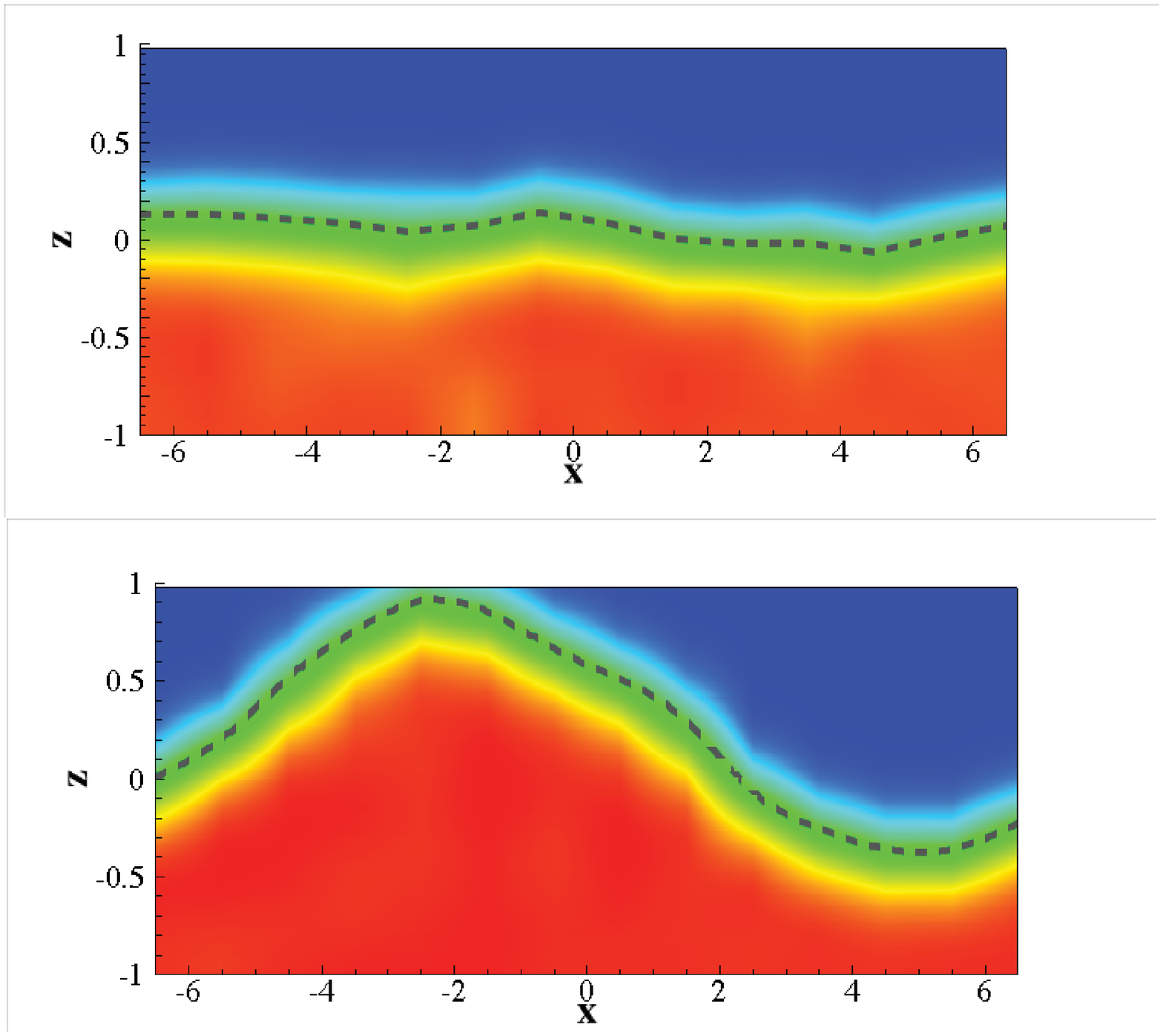}}
\caption{(a) Instantaneous fluid height near the interface with
  $\rho_{\alpha} = 32.0$ and
  $\rho_{\beta} = 64.0$ at $k_BT = 0.01$.
(b) Instantaneous fluid height near the interface with
$\rho_{\alpha} = 64.0$ and $\rho_{\beta} = 32.0$ at $k_BT = 0.01$.
(c) The capillary wave spectra measured at different gravity.
(d) Instantaneous density field $n_{\alpha}$ near the fluctuation interface
for $g = 0.0075$ (upper) and $g = 0.014$ (lower). For $g = 0.0075$, the
two-phase fluid keeps stable fluctuation interface with spectrum shown
in (c). For $g = 0.014$, Rayleigh instability is established and accompanied
by giant fluctuation across the interface.}
\label{fig:CWT_gravity}
\end{figure}

Next, we examine the interfacial fluctuations with a non-zero
gravity field and $k_B T=0.01$. We consider
two cases: (\Rmnum{1})
$\rho_{\alpha} = 64.0$, $\rho_{\beta} = 32.0$, and $g = 0.04$. (\Rmnum{2})
$\rho_{\alpha} = 32.0$, $\rho_{\beta} = 64.0$, and $g = 7.5\times10^{-3}$.
Figures.~\ref{fig:CWT_gravity}(a) and  \ref{fig:CWT_gravity}(b) show the instantaneous
interface $g(x,y)$ for cases (\Rmnum{1}) and  (\Rmnum{2}), respectively.
For case (\Rmnum{1}), the lower fluid has larger mass
density than the upper fluid. Contributions from the change of gravity potential
in Eq.~(\ref{eq:potential_contribution}) are positive, leading
to dampened interfacial
fluctuations in Figure~\ref{fig:CWT_gravity}(a)  compared to Figure~\ref{fig:CWT_neutral}(b).
In contrast, in case (\Rmnum{2}), the gravity contribution from
Eq.~(\ref{eq:potential_contribution}) is negative, leading
to increased interfacial
fluctuations in Figure~\ref{fig:CWT_gravity}(b)  compared to Figure \ref{fig:CWT_neutral}(b).

Figure~\ref{fig:CWT_gravity}(c) shows $\hat{\eta}(\mathbf{q})$
for cases (\Rmnum{1}) and (\Rmnum{2}). Numerical results are in good agreement 
with the predictions from Eq.~(\ref{eq:CWT_gravity}) for 
$|\mathbf{q}| \le \frac{2\pi}{5h}$. At high wave numbers, $\hat{\eta}(\mathbf{q})$
deviates from Eq.~(\ref{eq:CWT_gravity}) in a manner similar to the neutral case in
Figure~\ref{fig:CWT_neutral}(c).
For case (\Rmnum{2}), we choose $g$ and $\sigma$, satisfying
\begin{equation}
\sigma q_0^2 > (\rho_{\beta} - \rho_{\alpha})g,
\label{eq:Ray_instability}
\end{equation}
where $q_0 = \frac{2\pi}{L}$ is the lowest wave number such that Rayleigh
instability cannot be established. In contrast, if Eq.~(\ref{eq:Ray_instability})
is violated (e.g., $g = 0.014$ and $\sigma=2.0$) Rayleigh instability will be
develop as shown in Figure~\ref{fig:CWT_gravity}(d).

\section{Discussion}
\label{sec:discussion}
In this study, we proposed the rescaled Pairwise-Force Smoothed Dissipative Particle 
Hydrodynamics (rPF-SDPD) method, a fully Lagrangian stochastic particle method designed to model 
mesoscopic multicomponent  immiscible flow with thermal fluctuations. In the rPF-SDPD model, the surface tension 
between different fluid components is modeled via pairwise interaction forces added 
to the SDPD momentum conservation equation similar to the PF-SPH model \cite{Tart-PFSPH}. 
In PF-SPH, a relationship between the surface tension 
and  pairwise-force parameters (similar to Eq.~(\ref{s11})) is derived under a locally flat interface assumption.
In this work, we  demonstrated that, under moderate thermal fluctuations, the modeled
surface tension deviates from the analytical result given by Eq.~(\ref{s11}) 
 and also depends on the model resolution. 
To accurately model fluids interfaces, we  derived a universal scaling relationship between model parameters (macroscopic surface tension in the absence of thermal fluctuations, temperature, model resolution, and pairwise interaction force)
and the surface tension.
To establish this relation, we constructed a coarse-grained Euler lattice
model by mapping the SDPD particles on a discrete
lattice based on a mean field theory. 
We demonstrated that the numerical results obtained from the present rPF-SDPD
model agree well with the theoretical prediction based on the scaling 
relationship with deviation less than $5\%$.

Furthermore, we demonstrated that the rPF-SDPD model  yields consistent thermodynamic 
properties of the bulk fluid under thermal fluctuations.
Moreover, it accurately captures the dynamic processes,
such as the bubble coalescence and the capillary
wave spectrum under external gravity fields. 
These results
suggest that the present method is wellsuited for a 
wide application on multiphase immiscible flow on the mesoscopic
scale where thermal fluctuations are pronounced, including nanoscale transport processes.

Finally, we observed that  for interfaces with radii of curvature 
less than $2h$, the surface tension 
decreases with decreasing radii of curvature. Similar results 
are experimentally observed for real fluids, where the surface tension 
shows dependence on the radii of curvature for the radii on the order of the molecular size.
We note that this length-scale-dependent surface tension
(presented in Section~\ref{sec:sigma_curvature}) raises some 
important issues that require future investigation.
In most mesoscopic numerical methods (e.g., see 
Refs.~\cite{Hu2006, Hu_JCP_2009, Chaudhri_Donev_PRE_2014}) 
for multiphase and multicomponent flows, the interfacial
energy is imposed as the interface area multiplied by a prescribed  
surface tension coefficient. The implicit assumption therein 
is that the surface tension is a macroscopic property independent
of local interface curvature, i.e., it remains constant as
the spatial resolution of the interface increases.
This assumption works well for most macroscopic (and many
mesoscopic) multiphase flow systems. It also  can be achieved
for the present method by choosing proper scaling parameters so
 Eq.~(\ref{eq:R_criteria}) is satisfied. However, additional consistency is
required when we consider a multiphase flow system on the nanoscale. At this scale, arbitrarily increasing model
resolution leads to numerical divergence of interfacial
fluctuations, i.e., $\sim \displaystyle{\lim_{{q_h} \to \infty} 
\int_{q_l}^{q_h} \frac{1}{q^2} 
d^2q = \infty}$. On this length scale,  surface tension
also depends on the local curvature 
\cite{Lum_Chandler_JPCB_1999, Huang_Chandler_JPCB_2001} with behavior 
similar to Figure~\ref{fig:droplet_surface_tension}. 
For such systems, accurate fluctuation 
hydrodynamics modeling requires introduction of molecular fidelity in the form of an
effective particle size and local compressibility, 
as discussed in Ref. \cite{Lei_Schenter_JCP_2015}. 
Such use of additional collective variables will be explored in future work.

\begin{acknowledgments}
This research was supported by the U.S. Department of Energy, Office of Science, Office of Advanced 
Scientific Computing Research as part of the Collaboratory on Mathematics for Mesoscopic 
Modeling of Materials (CM4) and the New Dimension Reduction Methods and Scalable Algorithms for Nonlinear Phenomena project. CJM is supported by the DOE Office of Basic Energy Sciences, Division of Chemical Sciences, Geosciences and Biosciences. Pacific Northwest National Laboratory is operated by 
Battelle for the DOE under Contract DE-AC05-76RL01830.
HL would like to thank Bin Zheng for helpful discussions. 
\end{acknowledgments}

\clearpage
\section*{References}

\bibliographystyle{apsrev}
\bibliography{main}

\end{document}